\documentclass{elsarticle}

\usepackage[reviewcopy]{adndt}%{adndt} at NSCL, {../adndt} at Gross
\usepackage{longtable}

\usepackage{amsmath}

\usepackage{amssymb}

\biboptions{square,sort&compress}
\bibpunct[]{[}{]}{,}{n}{}{;}
\begin{document}

\begin{frontmatter}

\journal{Atomic Data and Nuclear Data Tables}

%% Author, fill in article title here

\title{Discovery of Scandium, Titanium, Mercury, and Einsteinium Isotopes}

%% Fill in author list here
\author{D. Meierfrankenfeld}
\author{A. Bury}
\author{M. Thoennessen\corref{cor1}}\ead{thoennessen@nscl.msu.edu}

 \cortext[cor1]{Corresponding author.}

 \address{National Superconducting Cyclotron Laboratory and \\ Department of Physics and Astronomy, Michigan State University, \\ East Lansing, MI 48824, USA}

\begin{abstract}
Currently, twenty-three scandium, twenty-five titanium, forty mercury and seventeen einsteinium isotopes have been observed and the discovery of these isotopes is discussed here. For each isotope a brief synopsis of the first refereed publication, including the production and identification method, is presented.
\end{abstract}

\end{frontmatter}

%%% Keywords and subject classification are not used in ADNDT
%%%\begin{keywords}
%%%Insert list of keywords here.
%%%\end{keywords}

%%%\begin{subject}[Insert header for classifications]
%%%Use only if your journal has a subject classification requirement
%%%\end{subject}

%%% The table of contents should start a new page. This command will
%%% automatically pull all the unstarred \section, \subsection and
%%% \subsubsection titles into the Contents. Starred versions need to be
%%% done manually using
%%%      \addcontentsline{toc}{[[sub]sub]section}{Section title}
%%% at the correct place. Examples are given below.

%%% The lists of data figures and data tables are created automatically
%%% by the \listofDfigures and \listofDtables commands.

\newpage
\tableofcontents
%%\listofDfigures
\listofDtables

\vskip5pc

\section{Introduction}\label{s:intro}

The discovery of scandium, titanium, mercury, and einsteinium isotopes is discussed as part of the series summarizing the discovery of isotopes, beginning with the cerium isotopes in 2009 \cite{2009Gin01}. Guidelines for assigning credit for discovery are (1) clear identification, either through decay-curves and relationships to other known isotopes, particle or $\gamma$-ray spectra, or unique mass and Z-identification, and (2) publication of the discovery in a refereed journal. The authors and year of the first publication, the laboratory where the isotopes were produced as well as the production and identification methods are discussed. When appropriate, references to conference proceedings, internal reports, and theses are included. When a discovery includes a half-life measurement the measured value is compared to the currently adopted value taken from the NUBASE evaluation \cite{2003Aud01} which is based on the ENSDF database \cite{2008ENS01}. In cases where the reported half-life differed significantly from the adapted half-life (up to approximately a factor of two), we searched the subsequent literature for indications that the measurement was erroneous. If that was not the case we credited the authors with the discovery in spite of the inaccurate half-life.

\section{Discovery of $^{39-61}$Sc}

Twenty-three scandium isotopes from A = $39-61$ have been discovered so far; these include 1 stable, 6 proton-rich and 16 neutron-rich isotopes. Many more additional neutron-rich nuclei are predicted to be stable with respect to neutron-emission and could be observed in the future. The mass surface towards the neutron dripline (the delineation where the neutron separation energy is zero) becomes very shallow. Thus the exact prediction of the location of the dripline is difficult and can vary substantially among the different mass models. As one example for a mass model we selected the HFB-14 model which is based on the Hartree-Fock-Bogoliubov method with Skyrme forces and a $\delta$-function pairing force \cite{2007Gor01}. According to this model, $^{68}$Sc should be the last odd-odd particle stable neutron-rich nucleus while the odd-even particle stable neutron-rich nuclei should continue through $^{77}$Sc. The proton dripline has been reached and no more long-lived isotopes are expected to exist because $^{39}$Sc has been shown to be unbound by 580 keV \cite{1988Woo01}. About 12 isotopes have yet to be discovered corresponding to 35\% of all possible scandium isotopes.

\begin{figure}
	\centering
	\includegraphics[scale=.5]{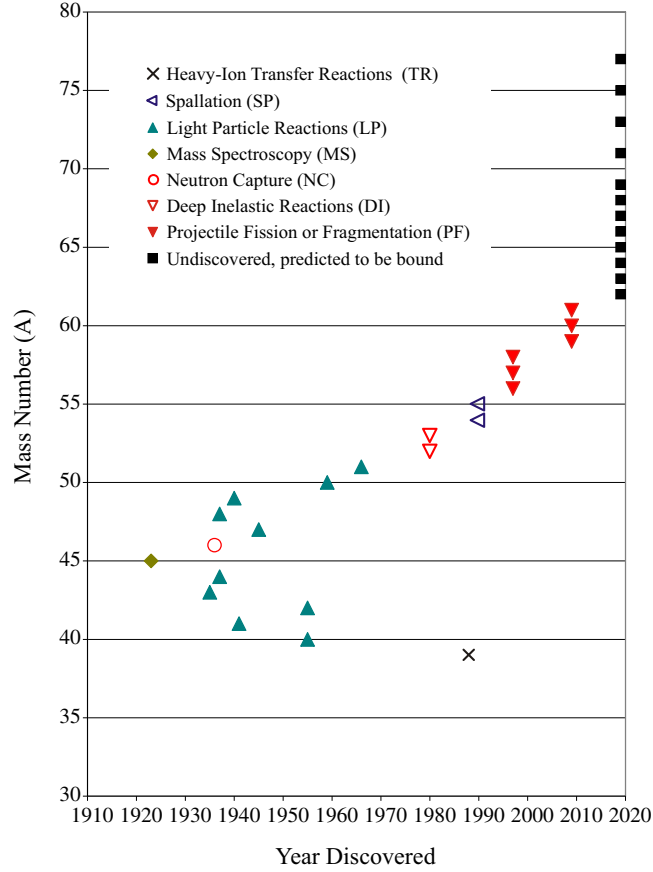}
	\caption{Scandium isotopes as a function of time when they were discovered. The different production methods are indicated. The solid black squares on the right hand side of the plot are isotopes predicted to be bound by the HFB-14 model.}
\label{f:year-sc}
\end{figure}

Figure \ref{f:year-sc} summarizes the year of first discovery for all scandium isotopes identified by the method of discovery. The range of isotopes predicted to exist is indicated on the right side of the figure. The radioactive scandium isotopes were produced using heavy-ion transfer reactions (TR), deep-inelastic reactions (DI), light-particle reactions (LP), neutron capture (NC), spallation (SP), and projectile fragmentation of fission (PF). The stable isotopes were identified using mass spectroscopy (MS). Heavy ions are all nuclei with an atomic mass larger than A=4 \cite{1977Gru01}. Light particles also include neutrons produced by accelerators. In the following, the discovery of each scandium isotope is discussed in detail and a summary is presented in Table 1.

\subsection{$^{39}$Sc}\vspace{0.0cm}

$^{39}$Sc was discovered by Woods \textit{et al.} in 1988 as reported in the paper ``A Measurement of the Mass of $^{39}$Sc'' \cite{1988Woo01}.  A beam of 102.5 MeV $^{14}$N accelerated by the 14UD pelletron accelerator at the Australian National University bombarded a $^{40}$Ca target on a carbon backing and $^{39}$Sc was identified by measuring the transfer reaction product $^{15}$C in an Enge split-pole spectrometer. ``A mass excess of $-$14.19$\pm$0.03 MeV has been derived for $^{39}$Sc from a measurement of the Q-value of the $^{40}$Ca($^{14}$N,$^{15}$C)$^{39}$Sc reaction.'' The observation of $^{39}$Sc was independently submitted a month later by Mohar \textit{et al.} \cite{1988Moh01}.

\subsection{$^{40}$Sc}\vspace{0.0cm}

Glass and Richardson discovered $^{40}$Sc in the 1955 article ``Radionuclides Al$^{24}$, P$^{28}$, Cl$^{32}$, and Sc$^{40}$'' \cite{1955Gla01}. A 20-MeV proton beam from the UCLA 41-in FM cyclotron bombarded a calcium target. The isotope was formed in the charge exchange reaction $^{40}$Ca(p,n)$^{40}$Sc. Positron and $\gamma$-ray spectra were measured with a NaI crystal. ``Sc$^{40}$ has a half-life of 0.22$\pm$0.03 sec, threshold of 15.9$\pm$1.0 Mev, a 3.75$\pm$0.04 Mev gamma ray, and maximum positron energy of 9.0$\pm$0.4 Mev.'' This half-life is close to the currently adapted value of 182.3(7)~ms. A previously reported observation of $^{40}$Sc estimated a half-life of $\approx$0.35~s which is almost off by a factor of two and the identification was only suggested ``from a simple consideration of preferred reaction type[s] and estimates of threshold[s]'' \cite{1954Tyr01}.

\subsection{$^{41}$Sc}\vspace{0.0cm}

In 1941 the discovery of $^{41}$Sc was reported by Elliot and King in the paper ``Radionuclides $_{21}$Sc$^{41}$, $_{18}$Al$^{35}$, and $_{16}$S$^{31}$'' \cite{1941Ell01}. $^{41}$Sc was produced in the reaction $^{40}$Ca($d,n$) using 8~MeV deuterons from the Purdue University cyclotron \cite{1941Ell02}. The half-life and the energy spectrum of the positrons were measured. ``Because Sc$^{41}$ cannot be reached by any other of the usual nuclear reactions its identification is mostly one of elimination. All other probable Ca(d,$-$) reactions have been investigated carefully'' \cite{1941Ell02}. The measured half-life of 0.87(3)~s is close to the adapted value of 593.6(17)~ms. Previously, a half-life of 52(2)~h had been reported which was changed to 53(3)~m in a note added in proof \cite{1937Wal01}. The assignment was based on the assumption that $^{44}$Ca was the heaviest stable isotope.

\subsection{$^{42}$Sc}\vspace{0.0cm}

Morinaga identified $^{42}$Sc correctly for the first time in the publication ``New Radioactive Isotope Scandium-42'' in 1955 \cite{1955Mor01}. At Purdue University a potassium target was bombarded with 18~MeV $\alpha$ particles. Decay curves of positrons and the annihilation radiation were measured with an anthracene and NaI crystal, respectively. ``A strong activity with a half-life of 0.62$\pm$0.05 sec (error limit) was found.'' This half-life agrees with the presently accepted value of 681.3(7)~ms. Previously, half-lives of 4.1(1)~h \cite{1937Wal05} and 13.5(3)~d \cite{1940Wal01} had incorrectly been assigned to $^{42}$Sc.

\subsection{$^{43}$Sc}\vspace{0.0cm}

$^{43}$Sc was discovered in 1935 by Frisch as reported in the paper ``Induced Radioactivity of Fluorine and Calcium'' \cite{1935Fri01}. Alpha particles from a 600 mCi radon source were used to irradiate calcium and $^{43}$Sc was formed in the reaction $^{40}$Ca($\alpha$,p). ``From the great intensity, one may say that the effect is due to the main isotope of calcium, Ca$^{40}$. Capture of the alpha particle, with subsequent emission of a proton or neutron, would lead to the formation of Sc$^{43}$ or Ti$^{43}$, respectively. A chemical separation, kindly carried out by Prof. G. von Hevesy, showed that the active body follows the reactions of scandium. Therefore the 4.4 hours activity certainly corresponds to Sc$^{43}$.'' This half-life, quoted with a possible error of 10\% agrees with the currently adapted value of 3.891(12)~h.

\subsection{$^{44}$Sc}\vspace{0.0cm}

In 1937 Walke observed $^{44}$Sc as described in the paper ``The Induced Radioactivity of Calcium'' \cite{1937Wal01}. $^{44}$Sc was identified in the reaction $^{41}$K($\alpha$,n) by bombarding potassium fluoride with 11 MeV $\alpha$ particles from the Berkeley cyclotron as described in a note added in proof: ``By deflecting the emitted particles in a magnetic field it has been established that they are positrons. The decay curve shows the presence of two isotopes with half-lives of 4.1$\pm$0.2 hours and 52$\pm$2 hours. As the long period agrees with that observed in the scandium precipitate from calcium + deuterons it must be associated with Sc$^{44}$...'' This half-life agrees with the currently accepted values of 3.97(4)~h. In the main text of the paper $^{44}$Sc had been assigned a half-life of 53$\pm$3~m. A 3 hour half-life had been reported earlier but it was assigned only to either $^{42}$Sc or $^{44}$Sc \cite{1934Zyw01}.

\subsection{$^{45}$Sc}\vspace{0.0cm}

Aston reported the discovery of $^{45}$Sc in the 1923 paper ``Further Determinations of the Constitution of the Elements by the Method of Accelerated Anode Rays'' \cite{1923Ast01}. The mass determination was made using the method of accelerated anode rays with a spectrograph: ``Scandium was successfully attacked by the use of material kindly supplied by Prof. Urbain, of Paris. The only line obtained was at 45. It may be taken provisionally to be a simple element, but the effects are not strong enough to disprove the presence of small quantities of another constituent.''

\subsection{$^{46}$Sc}\vspace{0.0cm}

$^{46}$Sc was discovered in 1936 by Hevesy and Levi reported in the paper ``The action of neutrons on the rare earth elements'' \cite{1936Hev01}. Scandium was irradiated by a 200-300 mCi radon-beryllium source and the activity was measured following chemical separation. ``The activities are due to the formation of $^{46}_{21}$Sc and $^{42}_{19}$K, respectively; the reaction leading to these products are $^{45}_{21}$Sc + $^{1}_{0}$n = $^{46}_{21}$Sc and $^{45}_{21}$Sc + $^{1}_{0}$n = $^{42}_{19}$K + $^{4}_{2}\alpha$. The mass numbers occuring in these equations follows from the fact that scandium has only one stable isotope, $^{45}_{21}$Sc... The activity which cannot be separated from scandium is presumable due to $^{46}_{21}$Sc; most of this activity decays with a period of about two months but a small part does not decay appreciably within a year or two.'' The half-life is close to the present value of 83.79(4)~d.

\subsection{$^{47}$Sc}\vspace{0.0cm}

$^{47}$Sc was identified correctly for the first time by Hibdon and Pool in the 1945 paper ``Radioactive Scandium. II'' \cite{1945Hib01}. 20~MeV $\alpha$-particles, 10~MeV deuterons and 5~MeV protons were accelerated by the Ohio State University 42-inch cyclotron and decay curves were measured with a Wulf unifilar electrometer \cite{1945Hib02}. ``This radioactive isotope [Sc$^{47}$] has been reported to have a half-life of 2.62 days and to emit 1.1 Mev electrons. These observations are not confirmed. A new radioactive isotope has, however, been produced by bombarding calcium with alpha-particles and to some extent by bombarding calcium with deuteron and proton. It emits beta-rays of 0.46 Mev and has a half-life of 3.4 days. This assignment is made to Sc$^{47}$.'' This half-life agrees with the presently adopted value of 3.3492(6)~d. In addition to the mentioned incorrect observation published in 1940 \cite{1940Wal01}, a 1938 paper had tentatively assigned the half-life of 28~h measured by Pool et al. \cite{1937Poo01} to $^{47}$Sc \cite{1938Cor01}.

\subsection{$^{48}$Sc}\vspace{0.0cm}

In 1937 the observation of $^{48}$Sc was reported in the paper ``The Induced Radioactivity of Titanium and Vanadium'' by H. Walke \cite{1937Wal03}. 14-20 MeV neutrons from the bombardment of 5.5 MeV deuterons on lithium at the Berkeley cyclotron were used to activate titanium and vanadium targets. $^{48}$Sc was produced in the reactions $^{48}$Ti(n,p) and $^{51}$V(n,$\alpha$). Decay and absorption measurements were performed with a Lauritsen type quartz fiber electroscope. ``When bombarded with fast neutrons it seems likely that Ca$^{45}$, Sc$^{48}$ and Sc$^{46}$ are formed from titanium. Sc$^{48}$ is also produced by the transmutation V$^{51}$ + n$^1 \rightarrow$ Sc$^{48}$ + He$^4$, its half-life being 41$\pm$3 hours.'' This half-life agrees with the currently accepted value of 43.67(9)~h.

\subsection{$^{49}$Sc}\vspace{0.0cm}

In 1940 Walke correctly identified $^{49}$Sc in the paper ``The Radioactive Isotopes of Scandium and Their Properties'' \cite{1940Wal01}. $^{49}$Sc was produced by bombarding calcium with 8 MeV deuterons from the 37-inch Berkeley cyclotron. $^{49}$Ca was identified by measuring the decay and absorption with a Lauritsen quartz fiber electroscope. The assignment was confirmed in the $\beta$-decay of $^{49}$Ca formed by neutron capture of $^{48}$Ca and in the reaction $^{49}$Ti(n,p)$^{49}$Sc. ``The activity of half-life 53$\pm$3 min., now measured accurately as 57$\pm$2 min., produced by bombarding calcium with deuterons, which emits $\beta$-particles of energy 1.8$\pm$0.1 MeV previously assigned to Sc$^{41}$ is shown to be probably due to Sc$^{49}$.'' The half-life of 57(2)~m agrees with the currently accepted value of 57.2(2)~m. The mentioned 53(3)~m half-life was originally assigned to $^{44}$Sc but changed to $^{41}$Sc in a note added in proof \cite{1937Wal01} which was still incorrect.

\subsection{$^{50}$Sc}\vspace{0.0cm}

$^{50}$Sc was discovered in 1959 by Poularikas and Fink in the publication ``Absolute Activation Cross Sections for Reactions of Bismuth, Copper, Titanium, and Aluminum with 14.8-Mev Neutrons'' \cite{1959Pou01}. Deuterons from the University of Arkansas 400-kV Cockroft-Walton Accelerator produced 14.8 MeV monoenergetic neutrons via the reaction $^3$H(d,n)$^4$He and $^{50}$Sc was produced in the reaction $^{50}$Ti(n,p). Decay curves were measured with a beta-proportional counter. A half-life of 1.80(20)~m was quoted for $^{50}$Sc in a table. This half-life is consistent with the presently accepted value of 102.5(5)~s. A second half-life of 22(3)~m tentatively assigned to $^{50}$Sc could not be confirmed \cite{1963Kan02,1963Koe01}. Poularikas and Fink did not acknowledge a previous half-life measurement of 1.8(2)~m only reported at a conference \cite{1955Mor02} and in an internal report \cite{1956Mor02}.

\subsection{$^{51}$Sc}\vspace{0.0cm}

Erskine et al. reported the first measurement of $^{51}$Sc in the 1966 paper ``Energy Levels in Sc$^{49}$ from Ca$^{48}$(He$^3$,d)Sc$^{49}$ and Other Reactions Proceeding from Ca$^{48}$ \cite{1966Ers01}. Alpha-particles from the Argonne tandem Van de Graaff accelerator bombarded evaporated $^{48}$Ca targets on a carbon backing. Ground-state Q-values and excited states were measured using the Argonne broad-range magnetic spectrograph. ``The isotope Sc$^{51}$ is observed here for the first time in the study of the Ca$^{48}$($\alpha$,p)Sc$^{51}$ reaction. A ground-state Q-value of $-$5.860$\pm$0.020 MeV was measured''. The first decay measurement of $^{51}$Sc was submitted only five months later \cite{1966Biz01}.

\subsection{$^{52,53}$Sc}\vspace{0.0cm}

$^{52}$Sc and $^{53}$Sc were first observed by Breuer \textit{et al.} in 1980 as described in the paper ``Production of neutron-excess nuclei in $^{56}$Fe-induced reactions'' \cite{1980Bre01}. $^{56}$Fe ions were accelerated to 8.3 MeV/u by the Berkeley Laboratory SuperHILAC accelerator and bombarded self-supporting $^{238}$U targets. New isotopes were produced in deep-inelastic collisions and identified with a $\Delta$E-E time-of-flight semiconductor detector telescope: ``...the identification of seven new isotopes is reported: $^{52-53}$Sc, $^{54-55}$Ti, $^{56}$V, and $^{58-59}$Cr.'' 30$\pm$8 and 19$\pm$6 events of $^{52}$Sc and $^{53}$Sc were observed, respectively.

\subsection{$^{54,55}$Sc}\vspace{0.0cm}

$^{54}$Sc and  $^{55}$Sc were discovered in 1990 by Tu \textit{et al.} in the paper ``Direct mass measurement of the neutron-rich isotopes of chlorine through iron'' \cite{1990Tu01}. 800 MeV protons from the Los Alamos Meson Physics Facility LAMPF bombarded a $^{nat}$Th target and the isotopes were identified using the Time-of-Flight Isochronous (TOFI) spectrometer. The mass excesses for 29 neutron-rich isotopes from chlorine to iron (including $^{54}$Sc and  $^{55}$Sc) were measured for the first time and presented in a table.

\subsection{$^{56-58}$Sc}\vspace{0.0cm}

Bernas \textit{et al.} observed $^{56}$Sc, $^{57}$Sc, and $^{58}$Sc for the first time in 1997 as reported in their paper ``Discovery and cross-section measurement of 58 new fission products in projectile-fission of 750$\cdot$A MeV $^{238}$U'' \cite{1997Ber01}. Uranium ions were accelerated to 750 A$\cdot$MeV by the GSI UNILAC/SIS accelerator facility and bombarded a beryllium target. The isotopes produced in the projectile-fission reaction were separated using the fragment separator FRS and the nuclear charge Z for each was determined by the energy loss measurement in an ionization chamber. ``The mass identification was carried out by measuring the time of flight (TOF) and the magnetic rigidity B$\rho$ with an accuracy of 10$^{-4}$.''  68, 30 and 11 counts of $^{55}$Sc, $^{56}$Sc and $^{57}$Sc were observed, respectively.

\subsection{$^{59-61}$Sc}\vspace{0.0cm}

$^{59}$Sc, $^{60}$Sc and $^{61}$Sc were observed by Tarasov et al. in 2009 and described in the publication ``Evidence for a change in the nuclear mass surface with the discovery of the most neutron-rich nuclei with 17 $\le$ Z $\le$ 25'' \cite{2009Tar01}. $^9$Be targets were bombarded with 132 MeV/u $^{76}$Ge ions accelerated by the Coupled Cyclotron Facility at the National Superconducting Cyclotron Laboratory at Michigan State University. $^{58}$Sc, $^{59}$Sc and $^{60}$Sc were produced in projectile fragmentation reactions and identified with a two-stage separator consisting of the A1900 fragment separator and the S800 analysis beam line. ``The observed fragments include fifteen new isotopes that are the most neutron-rich nuclides of the elements chlorine to manganese ($^{50}$Cl, $^{53}$Ar, $^{55,56}$K, $^{57,58}$Ca, $^{59,60,61}$Sc, $^{62,63}$Ti, $^{65,66}$V, $^{68}$Cr, $^{70}$Mn).''

\section{Discovery of $^{39-63}$Ti}

Twenty-five titanium isotopes from A = $39-63$ have been discovered so far; these include 5 stable, 7 proton-rich and 13 neutron-rich isotopes.  According to the HFB-14 model \cite{2007Gor01}, $^{78}$Ti should be the last even-even particle stable neutron-rich nucleus while the odd-even particle stable neutron-rich nuclei should continue through $^{69}$Ti. The proton dripline has been reached and no more long-lived isotopes are expected to exist because $^{38}$Ti has been shown to be unbound with an upper limit for the half-life of 120~ns \cite{1996Bla01}. About 12 isotopes have yet to be discovered. Almost 70\% of all possible titanium isotopes have been produced and identified so far.

\begin{figure}
	\centering
	\includegraphics[scale=.5]{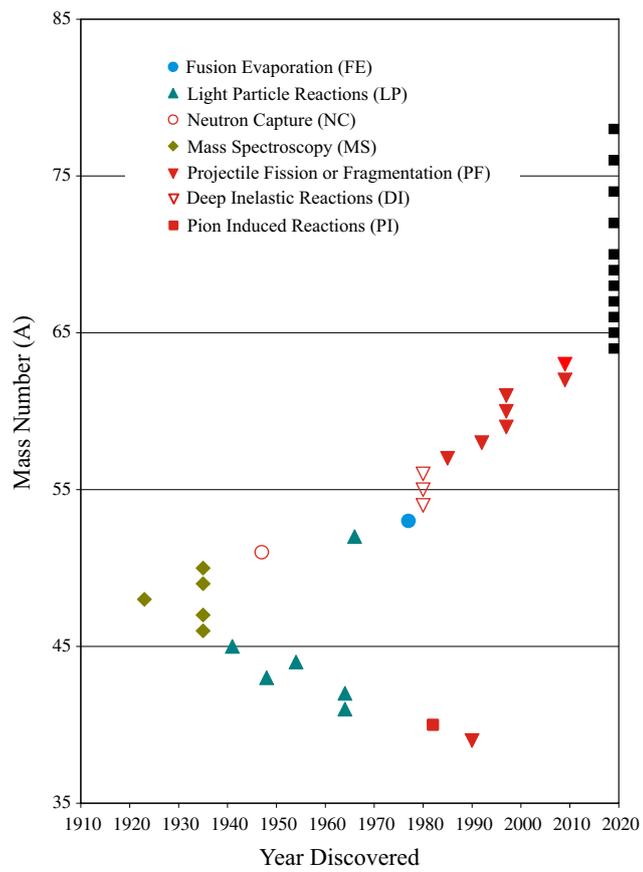}
	\caption{Titanium isotopes as a function of time when they were discovered. The different production methods are indicated. The solid black squares on the right hand side of the plot are isotopes predicted to be bound by the HFB-14 model.}
\label{f:year-ti}
\end{figure}

Figure \ref{f:year-ti} summarizes the year of first discovery for all titanium isotopes identified by the method of discovery.  The range of isotopes predicted to exist is indicated on the right side of the figure.  The radioactive titanium isotopes were produced using heavy-ion fusion evaporation (FE), deep-inelastic reactions (DI), light-particle reactions (LP), neutron-capture (NC), projectile fragmentation of fission (PF), and pion-induced reactions (PI). Heavy ions are all nuclei with an atomic mass larger than A=4 \cite{1977Gru01}. Light particles also include neutrons produced by accelerators. In the following, the discovery of each titanium isotope is discussed in detail and a summary is presented in Table 1.

\subsection{$^{39}$Ti}\vspace{0.0cm}

$^{39}$Ti was discovered by D\'{e}traz et al. in 1990, and published in ``Search for Direct Two-Proton Radioactivity from Ti Isotopes at the Proton Drip Line'' \cite{1990Det01}. $^{39}$Ti was produced by fragmenting $^{58}$Ni with the Grand Acc\'el\'rateur National d'Ions Lourds at Caen, France. The isotope was separated with the LISE spectrometer and identified by the observation of $\beta$-delayed particle emission. ``About 190 $^{40}$Ti ions and 75 $^{39}$Ti ions were collected during beam times of 3 and 24 hours, respectively.'' The measured half-life of 26$^{+8}_{-7}$~ms agrees with the adapted value of 31$^{+6}_{-4}$~ms.

\subsection{$^{40}$Ti}\vspace{0.0cm}

Morris et al. discovered $^{40}$Ti in ``Target mass dependence of isotensor double charge exchange: Evidence for deltas in nuclei'' in 1982 \cite{1982Mor01}. $^{40}$Ti was produced by the pion induced double charge exchange reaction $^{40}$Ca($\pi ^+$,$\pi ^-$) and the negative pions were anlysed with the  Energetic Pion Channel and Spectrometer EPICS. ``Byproducts of the present measurements are values of the masses of $^{28}$S and $^{40}$Ti. Our measured mass excesses are 4.13$\pm$0.16 and $-$8.79$\pm$0.16 MeV for $^{28}$S and $^{40}$Ti, respectively.''

\subsection{$^{41}$Ti}\vspace{0.0cm}

In 1964 Reeder et al. reported the discovery of $^{41}$Ti in ``New Delayed-Photon Emitters: Ti$^{41}$, Ca$^{37}$, and Ar$^{33}$'' \cite{1964Ree01}. Solid calcium targets were bombarded with 31.8~MeV $^3$He ions from the Brookhaven 60-inch cyclotron and $^{41}$Ti was produced in the reaction $^{40}$Ca($^3$He,2n). Surface-barrier detectors recorded proton spectra as a function of time. ``Three new nuclides, Ti$^{41}$, Ca$^{37}$, and Ar$^{33}$, have been observed to be delayed proton emitters.'' The measured half-life of 90.5(20)~ms agrees with the adopted value of 80.4(9)~ms.

\subsection{$^{42}$Ti}\vspace{0.0cm}

Bryant et al. discovered $^{42}$Ti in ``(He$^{3}$,n) Reactions on Various Light Nuclei'' in 1964 \cite{1964Bry01}. 25 MeV $^3$He ions from the Los Alamos variable energy cyclotron bombarded a solid calcium target and the excitation energy spectrum of $^{42}$Ti was extracted by measuring neutron energies with a bubble chamber. ``The neutron spectrum from [the $^{40}$Ca($^3$He,$n$)$^{42}$Ti] reaction is particularly interesting because, to our knowledge, the Ti$^{42}$ nucleus has not been previously studied. The energy of the neutrons corresponding to the ground state was estimated from the mass defect of $-$25.20 MeV (C$^{12}$ scale) predicted for Ti$^{42}$ ... which leads to a $Q$ for the above reaction of $-$2.79 MeV.''

\subsection{$^{43}$Ti}\vspace{0.0cm}

The discovery of $^{43}$Ti was reported in 1948 by Schelberg et al. in ``A Method for Measuring Short Period Activities'' \cite{1948Sch01}. 23 MeV $\alpha$ particles accelerated by the Indiana University 45-inch cyclotron bombarded metallic calcium and $^{43}$Ti was produced in the reaction $^{40}$Ca($\alpha$,n)$^{43}$Ti reaction. The pulses of a Geiger counter were stored in an oscilloscope and recorded by photographing the cathode-ray tube of the oscilloscope. ``The result obtained for Ti$^{43}$ is thus in quite good agreement with the value to be expected according to the analysis of Konopinski and Dickson.'' The measured half-life of 0.58(4)~s is consistent with the adopted value of 0.509(5)~s.

\subsection{$^{44}$Ti}\vspace{0.0cm}

In the 1954 paper ``A New Titanium Nuclide: Ti$^{44}$'' Sharp and Diamond announced the first observation of $^{44}$Ti \cite{1954Sha01}. Scandium oxide was bombarded with 30-45~MeV protons from the Harvard 95-inch synchrocyclotron. Activities were measured with a Geiger counter, a proportional counter and a NaI scintillation counter following chemical separation. ``Decay measurements made over a period of one half-year indicate, by a least-squares analysis, a half-life of 2.7~years with a rather large estimated error of $\pm$0.7~years because of the small amount of titanium activity available.'' This value was later changed to a half-life of larger than 23 years in an erratum \cite{1954Sha02}. This is consistent with the presently accepted value of 60.0(11)~y.

\subsection{$^{45}$Ti}\vspace{0.0cm}

$^{45}$Ti was discovered by Allen et al. in ``Artificial Radioactivity of Ti$^{45}$'' in 1941 \cite{1941All01}. $^{45}$Ti was produced by bombarding scandium oxide with 5 MeV protons from the Ohio State University 42-inch cyclotron. A Wulf electrometer with an ionization chamber recorded activities following chemical separation. ``The decay has been followed for more than eleven half-lives to an intensity of one-fourth background. The period obtained from this curve is 3.02 hours. Values determined for other samples bombarded under similar conditions are 3.17 and 3.10 hours... Since scandium has but a single stable isotope, the reaction Sc$^{45}$(p,n)Ti$^{45}$ should be the first considered.'' The measured half-life of 3.08(6)~h agrees with the adopted value of 3.08(1)~h.

\subsection{$^{46,47}$Ti}\vspace{0.0cm}

$^{46}$Ti and $^{47}$Ti were discovered by Aston in 1935 and published in ``The Isotopic Constitution and Atomic Weights of Hafnium, Thorium, Rhodium, Titanium, Zirconium, Calcium, Gallium, Silver, Carbon, Nickel, Cadmium, Iron and Indium'' \cite{1934Ast02}. The mass determination was made using a spectrograph with a discharge in titanium fluoride. ``The strong line 48 was found flanked by weak symmetrical pairs of satellites 46, 47, 49, 50, the whole forming a group of striking appearance.''

\subsection{$^{48}$Ti}\vspace{0.0cm}

Aston reported the discovery of $^{48}$Ti in the 1923 paper ``Further Determinations of the Constitution of the Elements by the Method of Accelerated Anode Rays'' \cite{1923Ast01}. The mass determination was made using a spectrograph. ``Titanium gives a strong line at 48.''

\subsection{$^{49,50}$Ti}\vspace{0.0cm}

$^{49}$Ti and $^{50}$Ti were discovered by Aston in 1935 and published in ``The Isotopic Constitution and Atomic Weights of Hafnium, Thorium, Rhodium, Titanium, Zirconium, Calcium, Gallium, Silver, Carbon, Nickel, Cadmium, Iron and Indium'' \cite{1934Ast02}. The mass determination was made using a spectrograph with a discharge in titanium fluoride. ``The strong line 48 was found flanked by weak symmetrical pairs of satellites 46, 47, 49, 50, the whole forming a group of striking appearance.''

\subsection{$^{51}$Ti}\vspace{0.0cm}

The first correct identification of $^{51}$Ti was reported in 1947 by Seren \text{et al.} in ``Thermal Neutron Activation Cross Sections'' \cite{1947Ser01}. Titanium metal powder was irradiated with thermal neutrons in the Argonne graphite pile reactor. Decay curves, and $\gamma$- and $\beta$-rays were measured. The observation of $^{51}$Ti is not specifically discussed among the 65 elements studied except in the main table: ``half-life previously reported 2.8 min. We find 6 min. over 3 half lifes.'' This half-life agrees with the presently accepted value of 5.76(1)~m. The quoted half-life of 2.8~m had been reported earlier \cite{1937Wal01} and could not be confirmed.

\subsection{$^{52}$Ti}\vspace{0.0cm}

Williams et al. discovered $^{52}$Ti in ``The (t,p) and (t,$\alpha$) Reactions on $^{48}$Ca and $^{50}$Ti'' in 1966 \cite{1966Wil01}. Tritons were accelerated to 7.5 MeV by the Los Alamos Van de Graaff Accelerator and bombarded a $^{50}$Ti target. $^{52}$Ti was produced in the (t,p) reaction and identified by measuring protons in a E-$\Delta$E solid-state detector system. ``For the $^{50}$Ti($t,p$)$^{52}$Ti reaction, Q$_0$ was found to be 5.698 $\pm$ 0.010 MeV.'' An earlier report of a 49(3)~m half-life \cite{1966Fac01} has not been confirmed.

\subsection{$^{53}$Ti}\vspace{0.0cm}

$^{53}$Ti was discovered by Parks et al. in 1977 and published in ``$\beta$ decay and mass of the new neutron-rich isotope $^{53}$Ti'' \cite{1977Par01}.$^{53}$Ti was produced in the fusion evaporation reaction $^{48}$Ca($^7$Li,pn) reaction at the Argonne FN tandem Van de Graaff accelerator. $\beta$ rays and $\beta$-delayed $\gamma$-ray were measured and ``The half-life of the decay of $^{53}$Ti was determined by following the decays of the $\beta$-delayed 101-, 127-, and 228-keV $\gamma$ rays. After correction for dead time, the composite decay curve for these three $\gamma$ rays yielded a half-life of 32.7$\pm$0.9 s.'' This half-life measurement is currently the only one for $^{53}$Ti.

\subsection{$^{54,55}$Ti}\vspace{0.0cm}

Guerreau et al. reported the discovery of $^{54}$Ti and $^{55}$Ti in the 1980 paper ``Seven New Neutron Rich Nuclides Observed in Deep Inelastic Collisions of 340 MeV $^{40}$Ar on $^{238}$U'' \cite{1980Gue01}. A 340 MeV $^{40}$Ar beam accelerated by the Orsay ALICE accelerator facility bombarded a 1.2 mg/cm$^2$ thick UF$_4$ target supported by an aluminum foil. The isotopes were identified using two $\Delta$E-E telescopes and two time of flight measurements. ``The new nuclides $^{54}$Ti, $^{56}$V, $^{58-59}$Cr, $^{61}$Mn, $^{63-64}$Fe, have been produced through $^{40}$Ar + $^{238}$U reactions.'' At least twenty counts were recorded for $^{54}$Ti. The identification of $^{55}$Ti was only tentative. Breuer et al. detected both isotopes independently only a few months later \cite{1980Bre01}.

\subsection{$^{56}$Ti}\vspace{0.0cm}

$^{56}$Ti was first observed by Breuer et al. in 1980 as described in ``Production of neutron-excess nuclei in $^{56}$Fe-induced reactions'' \cite{1980Bre01}. $^{56}$Fe ions were accelerated to 8.3 MeV/u by the Berkeley Laboratory SuperHILAC accelerator and bombarded self-supporting $^{238}$U targets. New isotopes were produced in deep-inelastic collisions and identified with a $\Delta$E-E time-of-flight semiconductor detector telescope. ``In addition, tentative evidence is found for $^{56}$Ti, $^{57-58}$V, $^{60}$Cr, $^{61}$Mn, and $^{63}$Fe.'' 13$\pm$5 events of $^{56}$Ti were observed.

\subsection{$^{57}$Ti}\vspace{0.0cm}

Guillemaud-Mueller et al. announced the discovery of $^{57}$Ti in the 1985 article ``Production and Identification of New Neutron-Rich Fragments from 33~MeV/u $^{86}$Kr Beam in the 18$\leq$Z$\leq$27 Region'' \cite{1985Gui01}.  At GANIL in Caen, France, a 33~MeV/u $^{86}$Kr beam was fragmented and the fragments were separated by the LISE spectrometer.  ``Each particle is identified by an event-by-event analysis.  The mass A is determined from the total energy and the time of flight, and Z by the $\delta$E and E measurements... In addition to that are identified the following new isotopes:  $^{47}$Ar, $^{57}$Ti, $^{59,60}$V, $^{61,62}$Cr, $^{64,65}$Mn, $^{66,67,68}$Fe, $^{68,69,70}$Co.'' At least four counts of $^{57}$Ti were observed.

\subsection{$^{58}$Ti}\vspace{0.0cm}

In their paper ``New neutron-rich isotopes in the scandium-to-nickel region, produced by fragmentation of a 500 MeV/u $^{86}$Kr beam'', Weber et al. presented the first observation of $^{58}$Ti in 1992 at GSI \cite{1992Web01}. $^{58}$Ti was produced in the fragmentation reaction of a 500 A$\cdot$MeV $^{86}$Kr beam from the heavy-ion synchroton SIS on a beryllium target and separated with FRS fragment separator. ``The isotope identification was based on combining the values of B$\rho$, time of flight (TOF), and energy loss ($\triangle$E) that were measured for each ion passing through the FRS and its associated detector array... The results ... represent unambiguous evidence for the production of the very neutron-rich isotopes $^{58}$Ti, $^{61}$V, $^{63}$Cr, $^{66}$Mn, $^{69}$Fe, and $^{71}Co...$''  Eleven events of $^{58}$Ti were observed.

\subsection{$^{59-61}$Ti}\vspace{0.0cm}

Bernas et al. observed $^{59}$Ti, $^{60}$Ti, and $^{61}$Ti for the first time in 1997 as reported in their paper ``Discovery and cross-section measurement of 58 new fission products in projectile-fission of 750$\cdot$A MeV $^{238}$U'' \cite{1997Ber01}. Uranium ions were accelerated to 750 A$\cdot$MeV by the GSI UNILAC/SIS accelerator facility and bombarded a beryllium target. The isotopes produced in the projectile-fission reaction were separated using the fragment separator FRS and the nuclear charge Z for each was determined by the energy loss measurement in an ionization chamber. ``The mass identification was carried out by measuring the time of flight (TOF) and the magnetic rigidity B$\rho$ with an accuracy of 10$^{-4}$.''  115, 40 and 9 counts of $^{59}$Ti, $^{60}$Ti and $^{61}$Ti were observed, respectively.

\subsection{$^{62,63}$Ti}\vspace{0.0cm}

$^{62}$Ti and $^{63}$Ti were discovered by Tarasov et al. in 2009 and published in ``Evidence for a change in the nuclear mass surface with the discovery of the most neutron-rich nuclei with 17 $\le$ Z $\le$ 25'' \cite{2009Tar01}. $^9$Be targets were bombarded with 132 MeV/u $^{76}$Ge ions accelerated by the Coupled Cyclotron Facility at the National Superconducting Cyclotron Laboratory at Michigan State University. $^{62}$Ti and $^{63}$Ti were produced in projectile fragmentation reactions and identified with a two-stage separator consisting of the A1900 fragment separator and the S800 analysis beam line. ``The observed fragments include fifteen new isotopes that are the most neutron-rich nuclides of the elements chlorine to manganese ($^{50}$Cl, $^{53}$Ar, $^{55,56}$K, $^{57,58}$Ca, $^{59,60,61}$Sc, $^{62,63}$Ti, $^{65,66}$V, $^{68}$Cr, $^{70}$Mn).''

\section{Discovery of $^{171-210}$Hg}

Forty mercury isotopes from A = $171-210$ have been discovered so far; these include 7 stable, 26 proton-rich and 7 neutron-rich isotopes.  According to the HFB-14 model \cite{2007Gor01}, $^{268}$Hg should be the last bound neutron-rich nucleus ($^{265}$Hg is predicted to be unbound). Along the proton dripline two more isotopes are predicted to be stable and it is estimated that seven additional nuclei beyond the proton dripline could live long enough to be observed \cite{2004Tho01}. Thus, there remain 66 isotopes to be discovered. Less than 40\% of all possible mercury isotopes have been produced and identified so far.

\begin{figure}
	\centering
	\includegraphics[scale=.5]{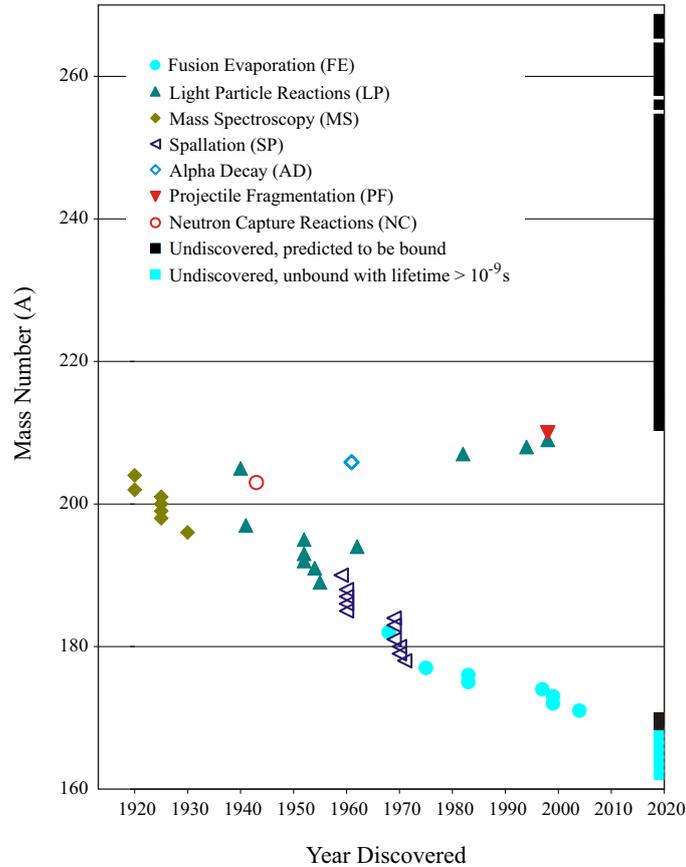}
	\caption{Mercury isotopes as a function of time they were discovered. The different production methods are indicated. The solid black squares on the right hand side of the plot are isotopes predicted to be bound by the HFB-14 model.  On the proton-rich side the light blue squares correspond to unbound isotopes predicted to have lifetimes larger than $\sim 10^{-9}$~s.}
\label{f:year-hg}
\end{figure}

Figure \ref{f:year-hg} summarizes the year of first discovery for all mercury isotopes identified by the method of discovery.  The range of isotopes predicted to exist is indicated on the right side of the figure.  The radioactive mercury isotopes were produced using fusion evaporation (FE), light-particle reactions (LP), neutron-capture (NC), spallation reactions (SP), projectile fragmentation or fission (PF), and $\alpha$-decay (AD). The stable isotopes were identified using mass spectroscopy (MS). Heavy ions are all nuclei with an atomic mass larger than A=4 \cite{1977Gru01}. Light particles also include neutrons produced by accelerators. In the following, the discovery of each mercury isotope is discussed in detail and a summary is presented in Table 1.

\subsection{$^{171}$Hg}\vspace{0.0cm}
$^{171}$Hg was discovered by Kettunen et al. in 2004, and published in ``Decay studies of $^{170,171}$Au, $^{171-173}$Hg, and $^{176}$Tl'' \cite{2004Ket01}. The cyclotron of the Accelerator Laboratory of the University of Jyv\"askyl\"a in Finland was used to produce the isotopes by the $^{96}$Ru($^{78}$Kr,3n) fusion evaporation reaction. The isotopes were separated by the gas-filled recoil separator RITU, and implanted into a position sensitive silicon strip detector. The identification was based on correlations between the daughter and granddaughter activities of $^{167}$Pt and $^{163}$Os, respectively.``A new $\alpha$-decaying isotope $^{171}$Hg and the previously known $^{172}$Hg isotope were produced via 3n- and 2n-fusion evaporation channels in the bombardment of the $^{96}$Ru target with the $^{78}$Kr ion beam.'' The reported half-life of 59$^{+36}_{-16}$~$\mu$s is currently the only measurement.

\subsection{$^{172,173}$Hg}\vspace{0.0cm}

The discovery by Seweryniak et al. of $^{172}$Hg and $^{173}$Hg was published in ``Decay properties of the new isotopes $^{172}$Hg and $^{173}$Hg'' in 1999 \cite{1999Sew01}. The isotopes $^{172}$Hg and $^{173}$Hg were produced by the ATLAS superconducting linear accelerator at Argonne National Laboratory by the fusion evaporation reactions, $^{78}$Kr($^{96}$Ru,2n) and $^{80}$Kr($^{96}$Ru,3n), respectively. To identify the mass, the isotopes were separated using the Argonne Fragment Mass Analyzer. Three strong lines ``are also present in decays followed within 100 ms by $\alpha$ particles corresponding to the decay of $^{168}$Pt (E$_{\alpha} \approx $6.83~MeV), and thus are assigned to the decay of the previously unknown isotope $^{172}$Hg.'' The observation of $\alpha$-radiation populating $^{169}$Pt ``leads to an unambiguous assignment of this group to the decay of a new isotope $^{173}$Hg.'' The 0.25$^{+.35}_{-.09}$~ms measured half-life of $^{172}$Hg agrees with the adopted value of 0.29$^{+0.23}_{-0.08}$~ms. The reported half-life for $^{173}$Hg of 0.9$^{+0.6}_{-0.3}$~ms is the only measurement quoted by the ENSDF database, while the NUBASE evaluation lists a half-life of 1.1(4)~ms. More recently a half-life of 0.59$^{+0.47}_{-0.18}$~ms was measured \cite{2004Ket01}.

\subsection{$^{174}$Hg}\vspace{0.0cm}

Uusitalo et al. published the discovery of $^{174}$Hg in ``Alpha decay of the new isotope $^{174}$Hg'' \cite{1997Uus01} in 1997. $^{174}$Hg was produced in the fusion-evaporation reaction $^{144}$Sm($^{36}$Ar,6n) with beam energy between 180 and 230~MeV at the University of Jyv\"askyl\"a in Finland. The isotopes were separated by the gas-filled recoil separator RITU, and implanted into a position sensitive silicon strip detector. ``The measured decay properties of the daughter activity are compatible with those of $^{170}$Pt for which T$_{1/2}$ = 6$^{+5}_{-2}$~ms and E$_{\alpha}$ = 6545$\pm$8~keV were reported. The mother activity can then be assigned to the new mercury isotope $^{174}$Hg.'' The reported half-life of 2.1$^{+1.8}_{-0.7}$~ms agrees with the currently adopted value of 2.0(4)~ms.

\subsection{$^{175,176}$Hg}\vspace{0.0cm}

``Alpha Decay of New Neutron Deficient Gold, Mercury and Thallium Isotopes'' reported the discovery of $^{175}$Hg and $^{176}$Hg by Schneider et al. in 1983 at the Gesellschaft f\"{u}r Schwerionenforschung (GSI) in Germany \cite{1983Sch01}. The isotopes were produced in fusion-evaporation reactions of a $^{92}$Mo beam with energies between 4.5 A$\cdot$MeV and 5.4 A$\cdot$MeV and separated with the velocity filter SHIP. ``The decays of the new isotopes $^{173}$Au, $^{175,176}$Hg and $^{179}$Tl could be correlated to the known $\alpha$-decays of their daughters''. The half-life of $^{175}$Hg was measured to be 20$^{+40}_{ -13}$~ms, while the half life of $^{176}$Hg was measured to be 34$^{+18}_{-9}$~ms. Both half-lives are close to the adopted values of 10.8(4)~ms and 20.3(14)~ms for $^{175}$Hg and $^{176}$Hg, respectively.

\subsection{$^{177}$Hg}\vspace{0.0cm}

$^{177}$Hg was first observed in 1975 in ``Mise en \'{e}vidence d'un nouvel isotope de mercure de masse 177'' by Cabot et al. \cite{1975Cab02}. A neodymium target was bombarded by a beam of calcium ions from the ALICE accelerator at Orsay, and the isotope was created by the reaction $^{142}$Nd($^{40}$Ca,5n)$^{177}$Hg. ``L'identification de $^{177}$Hg nous paraît donc bien établie'' (The identification of $^{177}$Hg seems to us well established). No half-life measurement was made.

\subsection{$^{178}$Hg}\vspace{0.0cm}

In 1971 Hansen et al. discovered $^{178}$Hg in ``The Alpha Decay of $^{179}$Hg and $^{178}$Hg'' \cite{1971Han01}. A lead target was bombarded with 600~MeV protons by the synchrocyclotron at CERN. $^{178}$Hg was separated and identified with the the isotope-separator-on-line facility ISOLDE by measuring $\alpha$ energies and half-lives. ``The assignment of the 6425 keV line to $^{178}$Hg cannot be doubted since all conceivable contaminants have lower alpha energies." The measured half-life of 0.47(14)~s agrees with the currently adopted value of 0.266(25)~s.

\subsection{$^{179,180}$Hg}\vspace{0.0cm}

$^{179}$Hg and $^{180}$Hg were discovered by Hansen et al. in 1970 and reported in ``Studies of the $\alpha$-active isotopes of mercury, gold and platinum'' \cite{1970Han01}. A lead target was bombarded with 600~MeV protons by the synchrocyclotron at CERN and $^{179}$Hg and $^{180}$Hg were separated and identified with the the isotope-separator-on-line facility ISOLDE. ``The $\alpha$-decay energy for $^{180}$Hg is 6.118 MeV; T$_{1/2}$ was found to be 2.9$\pm$0.3 sec... The line at 6.270 MeV, containing altogether 19 counts, is from $^{179}$Hg. The assignment is certain, because the energy is higher than any other observed in connection with the study of the mercury isotope.'' The measured half-life of 2.9(3)~s for $^{180}$Hg agrees with the currently adopted value of 2.58(1)~s. The half-life of $^{179}$Hg was not determined.

\subsection{$^{181}$Hg}\vspace{0.0cm}

Hansen et al. reported the first observation of $^{181}$Hg in the paper ``Decay Characteristics of Short-Lived Radio-Nuclides Studied by On-Line Isotope Separator Techniques'' in 1969 \cite{1969Han01}. 600 MeV protons from the CERN synchrocyclotron bombarded a lead target and $^{181}$Hg was separated using the ISOLDE facility. The paper summarized the ISOLDE program and did not contain details about the individual nuclei other than in tabular form. The detailed analysis was published in reference \cite{1970Han01}. The measured half-life of 3.6(3)~s was misprinted as 3.6(3)~m in \cite{1969Han01} and corrected in an errata. This value agrees with the adopted value of 3.6(1)~s.

\subsection{$^{182}$Hg}\vspace{0.0cm}

$^{182}$Hg was discovered by Demin in 1968 et al. in ``New Mercury Isotopes'' \cite{1968Dem01}. $^{182}$Hg was produced by the fusion-evaporation reactions $^{147}$Sm + $^{40}$Ar and $^{170}$Yb + $^{20}$Ne at the homogeneous magnetic-field cyclotron of the Laboratory for Nuclear Reactions in Dubna, Russia. The alpha radiation was then analyzed by a semiconductor spectrometer. ``The $^{182}$Hg assignment is confirmed also by the fact that the decay curve of $^{178}$Pt in the appropriate part of the excitation function shows a slight activation in accordance with the half-life of $^{182}$Hg.'' The measured half life of 9.6(2)~s is close to the adopted value of 10.83(6)~s.

\subsection{$^{183,184}$Hg}\vspace{0.0cm}

Hansen et al. reported the first observations of $^{183}$Hg and $^{184}$Hg in the paper ``Decay Characteristics of Short-Lived Radio-Nuclides Studied by On-Line Isotope Separator Techniques'' in 1969 \cite{1969Han01}. 600 MeV protons from the CERN synchrocyclotron bombarded a lead target and $^{183}$Hg and $^{184}$Hg were separated using the ISOLDE facility. The paper summarized the ISOLDE program and did not contain details about the individual nuclei other than in tabular form. The detailed analysis was published in reference \cite{1970Han01}. The measured half-lives of 8.8(5)~s for $^{183}$Hg and 32.0(10)~s for $^{184}$Hg agree with the adopted values of 9.4(7)~s for $^{183}$Hg and 30.9(3)~s for $^{184}$Hg. The previously reported half-life for $^{183}$Hg of approximately 26~s \cite{1968Dem01} was misidentified and may have been from the decay of $^{184}$Hg.

\subsection{$^{185-187}$Hg}\vspace{0.0cm}

$^{185}$Hg, $^{186}$Hg and $^{187}$Hg were first observed by Albouy et al. in 1960: ``Noveaux isotopes de p\'eriode courte obtenus par spallation de l'or'' \cite{1960Alb01}. Gold targets were bombarded with 155 MeV protons from the Orsay synchro-cyclotron and the isotopes were produced in spallation reactions. Half-life and $\gamma$-ray measurements were performed following double magnetic separation.``Les isotopes de mass 187, 186 et 185, de p\'eriode courte, ont pu \^etre observ\'es gr\^ace au montage d'un scintillateur \`a l'int\'erieur du s\'eparateur, derri\`ere le collecteur du 2$^e$ \'etage.'' (The short-lived isotopes of mass 187, 186, and 185 could be observed thanks to a scintillator mounted inside the separator after the collector of the second stage.) The measured half-lives of 50~s ($^{185}$Hg), 1.5~m ($^{186}$Hg) and 3~m ($^{187}$Hg) are consistent with the adopted values of 49.1(10)~s, 1.38(6)~min, and 2.4(3)~min, respectively.

\subsection{$^{188}$Hg}\vspace{0.0cm}

N. Poffe et al. reported the first observation of $^{188}$Hg in ``R\'eactions (p,xn) induites dans l'or par des protons de 155 MeV'' in 1960 \cite{1960Pof01}. Gold targets were bombarded with 155 MeV protons from the synchrocyclotron of the Paris Faculty of Sciences and $^{188}$Hg was identified by half-life and $\gamma$-ray measurements following magnetic separation: ``Half-lives and main $\gamma$-ray energies have been measured for $^{190}$Hg, $^{189}$Hg, $^{188}$Hg and their daughter products.'' The quoted half-life of 3.7~m is close to the currently accepted value of 3.25(3)~min.

\subsection{$^{189}$Hg}\vspace{0.0cm}

Smith and Hollander first observed $^{189}$Hg in 1955 and the results were reported in ``Radiochemical Study of Neutron-Deficient Chains in the Noble Metal Region'' \cite{1955Smi01}. A gold target was bombarded with a beam of 120 MeV protons accelerated with the Berkeley 184-cyclotron producing $^{189}$Hg in (p,9n) reactions. Identification was achieved by timed chemical separation of the $^{189}$Au daughter nuclei. ``The experiments cited here which establish the 42-minute gold as Au$^{189}$ therefore also set the half-life of Hg$^{189}$ as approximately 20 minutes.'' This 20(10)~m half-life is consistent with the present value of 7.6(1)~m for the ground state or the 8.6(1)~m isomeric state.

\subsection{$^{190}$Hg}\vspace{0.0cm}

G. Albouy et al. observed $^{190}$Hg for the first time in 1959: ``Isotopes de nombe de masse 190 du mercure et de l'or'' \cite{1959Alb01}. Mercury isotopes were produced in (p,xn) spallation reactions at the Orsay synchrocyclotron and mass 190 was selected with an isotope separator. K-X-rays were detected with a NaI(Tl) detector. ``Nous avons observ\'e dans la d\'ecroissance du rayonnement K principalement deux p\'eriodes en filiation 21$\pm$2 mn suivie de 45$\pm$3 mn. Nous attribuons ces p\'eriodes respectivement \'a $^{190}$Hg et \'a son descendant $^{190}$Au.'' (We observed the decay by K-radiation to consist mainly of two correlated periods, 21$\pm$2~m followed by 45$\pm$3~m. We attributed these periods with $^{190}$Hg and its daughter $^{190}$Au, respectively.) The half-life for $^{190}$Hg agrees with the accepted value of 20.0(5)~m.

\subsection{$^{191}$Hg}\vspace{0.0cm}

$^{191}$Hg was first identified by Gillon et al. from Princeton University in ``Nuclear Spectroscopy of Neutron-Deficient Hg Isotopes'' in 1954 \cite{1954Gil01}. The experiment was performed at the Harvard Cyclotron Laboratory using the reaction $^{197}$Au(p,7n) at 65 MeV. Conversion electron spectra were measured with a 119-gauss magnet and K-X-ray lines were attributed to $^{191}$Au following the decay of $^{191}$Hg. ``A set of lines decaying with half life 57(5)~min was observed'' This half-life agrees with the 49(10)~m half-life of the ground-state as well as with the 50.8(15)m half-life of the isomeric state.

\subsection{$^{192,193}$Hg}\vspace{0.0cm}

The discovery of $^{192}$Hg and $^{193}$Hg was published in 1952 by Fink et al. in ``Neutron-deficient Mercury Isotopes'' \cite{1952Fin01}. The Rochester cyclotron was used to bombard gold with 55, 65, and 96 MeV protons. $^{192}$Hg was identified by the activity from the $^{192}$Au daughter. ``With protons of 60 to 96 mev. on 0.003 inch gold foil in the Rochester cyclotron, a new mercury activity of half-life 5.7 $\pm$ 0.5 hours is observed.'' For $^{193}$Hg, ``the requisite quantitative relationship between 10-hour mercury and 15.3 hour Au$^{193}$ daughter has been demonstrated.'' The half-life of $^{192}$Hg is consistent with the adopted value of 4.85(20)~hours, while the 10.0(5)~hour measured half-life of $^{193}$Hg most probably corresponds to the 11.8(2)~hour isomeric state.

\subsection{$^{194}$Hg}\vspace{0.0cm}

$^{194}$Hg was first observed in 1962 by Tomlinson et al. in ``Nuclear Isomer Shift in the Optical Spectrum of Hg$^{195}$: Interpretation of the Odd-Even Staggering Effect in Isotope Shift'' \cite{1962Tom01}. $^{194}$Hg was produced in the reaction $^{197}$Au(p,4n) reaction by bombarding a gold target with 30~MeV protons from the Harvard University Cyclotron. The isotope shift of $^{194}$Hg was measured with a mirror monochromator M.I.T. after the other radioactive mercury isotopes had decayed \cite{1964Tom01}. ``In Table I we show the results for the isotope shifts in the 2537~$\AA$ line. These include the present work on Hg$^{194}$, Hg$^{195}$, and Hg$^{195m}$ ...'' The isotope was assumed to be known, however, the half-life measurements were incorrect \cite{1955Bru01,1961Mer01}.

\subsection{$^{195}$Hg}\vspace{0.0cm}

The discovery of $^{195}$Hg was published by Fink et al. in 1952 in ``Neutron-deficient Mercury Isotopes'' \cite{1952Fin01}. The Berkeley 184-inch cyclotron was used to bombard gold with 30 to 60~MeV protons. Half-life measurements were performed following chemical separation: ``A 31-hour activity from 30 mev. proton bombardment of gold is assigned to Hg$^{195}$.'' This half-life is most probably a measurement of the 41.6(3)~h isomeric state.

\subsection{$^{196}$Hg}\vspace{0.0cm}

Aston first identified $^{196}$Hg in ``Photometry of Mass-Spectra'' in 1930 \cite{1930Ast01}. The isotope was discovered by taking a mass-spectrograph of naturally occurring mercury. ``The rough value for the very faint isotope 196 was obtained as in the case of xenon by contour plotting from a very long exposure.''

\subsection{$^{197}$Hg}\vspace{0.0cm}

$^{197}$Hg was first identified in 1941 by Sherr et al. in ``Transmutation of Mercury by Fast Neutrons'' \cite{1941She01}. A mercury target was bombarded with fast neutrons from the Li + d reaction. The deuterons were produced by the Harvard cyclotron. Referring to a private communication by G.E. Valley the paper states: ``The assignment of the 25-hour period to Hg$^{197}$ is consistent with the present investigations.'' This half-life agrees with the adopted value of 23.8(1)~h. Previously a $\sim$ 45-m half-life first observed by Heyn \cite{1937Hey01} was incorrectly assigned to $^{197}$Hg \cite{1938Alv02}.

\subsection{$^{198-201}$Hg}\vspace{0.0cm}
The stable isotopes $^{198-201}$Hg were discovered by Aston in 1925 published in {\it{The Isotopes of Mercury}} \cite{1925Ast01}. A new mass spectrograph with twice the dispersion of the previous one made the identification of the mercury isotopes possible. ``Preliminary photographs of the mass-spectra of mercury show its lines clearly resolved and so enable a definite statement to be made on the mass numbers of its most important constituents.''

\subsection{$^{202}$Hg}\vspace{0.0cm}

In 1920 Aston reported the discovery of the stable isotope $^{202}$Hg in {\it{The Constitution of the Elements}} \cite{1920Ast01}. Aston used a mass spectrograph to identify two isotopes of mercury: ``Further examination of the multiply charged mercury clusters indicate the probability of a strong line at 202, a weak component at 204.'' Aston published the actual spectra a few months later \cite{1920Ast03}.

\subsection{$^{203}$Hg}\vspace{0.0cm}
$^{203}$Hg was first identified by Friedlander et al. in 1943 in ``Radioactive Isotopes of Mercury'' \cite{1943Fri01}. The isotope was formed by irradiating stable mercury isotopes with neutrons produced by the bombardment of lithium with 14~MeV deuterons from the 60-inch cyclotron at the University of California, Berkeley. ``Since both slow and fast neutrons produce the activity, the best assignment is Hg$^{203}$ which can be produced by $n$-$\gamma$ reaction from Hg$^{202}$, and by $n$-$2n$ reaction from Hg$^{204}$.'' The half-life of 51.5(15)~d is consistent with the adopted value of 46.594(12)~d. In 1937 a $\sim$45~m half-life had been incorrectly assigned to $^{203}$Hg \cite{1937McM01}. A half-life of $\sim$50~d had been previously observed, but no mass assignment was made \cite{1941She01}.

\subsection{$^{204}$Hg}\vspace{0.0cm}

In 1920 Aston reported the discovery of the stable isotope $^{204}$Hg in {\it{The Constitution of the Elements}} \cite{1920Ast01}. Aston used a mass spectrograph to identify two isotopes of mercury: ``Further examination of the multiply charged mercury clusters indicate the probability of a strong line at 202, a weak component at 204.'' Aston published the actual spectra a few months later \cite{1920Ast03}.

\subsection{$^{205}$Hg}\vspace{0.0cm}
$^{205}$Hg was discovered by Krishnan et al. in 1940 and published in ``Deuteron Bombardment of the Heavy Elements'' \cite{1940Kri01}. Mercury was bombarded with 9~MeV deuteron beams from the Cavendish cyclotron. Resulting beta radiation ``has been assigned to Hg$^{205}$ decaying to Tl$^{205}$ by the emission of continuous $\beta$-rays.'' The measured half-life of 5.5(2)~m agrees with the adopted value of 5.14(9)~m. Previously a half-life measurement of 40~h half-life was incorrectly assigned to $^{205}$Hg \cite{1936And02}.

\subsection{$^{206}$Hg}\vspace{0.0cm}
$^{206}$Hg was discovered by Nurmia et al. in 1961 and published in ``Mercury-206: a New Natural Radionuclide'' \cite{1961Nur01}. The isotope was formed by the alpha decay of $^{210}$Pb at the Institute of Physics of the University of Helsinki, Finland. Beta-decay curves were recorded following chemical separation. ``The hitherto unknown mercury-206 is formed from lead-210 by alpha decay.'' The measured half-life of 7.5(10)~min agrees with the adopted value of 8.32(7)~min.

\subsection{$^{207}$Hg}
Mirzadeh et al. discovered $^{207}$Hg in 1982 in ``A Rapid Radiochemical Separation Procedure for Mercury from Lead and Bismuth Targets'' \cite{1982Mir01}. The isotope was formed by bombarding lead and bismuth with 30-160~MeV neutrons from the Brookhaven Medium Energy Intense Neutron facility. Gamma-ray spectra were measured following chemical separation. ``As part of the systematic studies of fast neutron cross sections for targets over the entire periodic table, we have produced $^{205-207}$Hg by $^{208}$Pb/n,2pxn/ and $^{209}$Bi/n,3pxn/ reactions.'' No half life measurement was made. Mirzadeh et al. mentions a previously measured half-life without a reference. It is interesting to note that presently the only measured half-life was only published in a 1981 conference proceedings \cite{1981Jon01}.

\subsection{$^{208}$Hg}
$^{208}$Hg was discovered by Zhang et al. in 1994 in ``Observation of the new neutron-rich nuclide $^{208}$Hg'' \cite{1994Zha01}. A lead target was bombarded by a 30~MeV/nucleon $^{12}$C beam from the Heavy Ion Research Facility Lanzhou at the Institute of Modern Physics, Lanzhou, China. ``The assignment of $^{208}$Hg was based on the identification of its $\beta ^-$ decay daughter $^{208}$Tl observed in the periodically extracted Tl element sample growing in the separated Hg element product solution.'' The measured half-life of 42$^{+23}_{-12}$~m agrees with the adopted value of 41$^{+5}_{-4}$~m.

\subsection{$^{209}$Hg}
In 1998 Zhang et al. reported the discovery of $^{209}$Hg in ``Neutron-rich heavy residues and exotic multinucleon transfer'' \cite{1998Zha01}. A lead target was bombarded by a 50~MeV/nucleon $^{18}$O beam from the Heavy Ion Research Facility at the Institute of Modern Physics in Lanzhou, China. The isotopes were separated online with a gas-thermochromatographic device and $\gamma$-$\beta$ coincidences were recorded. ``$^{209}$Hg was created through an exotic $-$2p3n multinucleon transfer process and was identified for the first time.'' The half-life value of 35$^{+9}_{-6}$~s is currently the only measurement. It should be noted that Zhang submitted the results simultaneously to a separate journal \cite{1998Zha02}.

\subsection{$^{210}$Hg}
$^{210}$Hg was produced by Pf\"{u}tzner et al. in 1998 in ``New isotopes and isomers produced by the fragmentation of $^{238}$U at 1000 MeV/nucleon'' \cite{1998Pfu01}. Projectile fragmentation was used to produce the isotope by bombarding a beryllium target with a 1000~MeV/nucleon $^{238}$U beam from the SIS/FRS facility at GSI in Darmstadt, Germany. ``Although only five counts represent the experimental evidence for the observation of $^{210}$Hg, the very small number of background events on the identification plots supports the adopted assignment.''

\section{Discovery of the Element Einsteinium}
Einsteinium is the first transuranium element covered in this series and thus it is appropriate to discuss the discovery of the element itself first. While the criteria for the discovery of an element are well established \cite{1976Har02,1990Sea01,1991IUP01} the criteria for the discovery or even the existence of an isotope are not well defined (see for example the discussion in reference \cite{2004Tho01}). Therefore it is possible, as in the present case of einsteinium, that the discovery of an element does not necessarily coincide with the first discovery of a specific isotope. A common criterium is the requirement that the discovery has to be published in a refereed journal.

The new element with Z = 99 was first identified by Ghiorso et al. on December 19-20, 1952 from uranium which had been irradiated by neutrons in the ``Mike'' thermonuclear explosion on November 1, 1952 \cite{1955Ghi01}. However, the work was classified and could not be published. The authors realized the possibility that others could produce this new element independently, publish the results first and take credit for the discovery: ``At this juncture we began to worry that other laboratories might discover lighter isotopes of the elements 99 and 100 by the use of reactions with cyclotron-produced heavy ions. They would be able to publish that work without any problem and would feel that they should be able to name these elements. This might well happen before we could declassify the Mike work and it would make it difficult for us to claim priority in discovery. (Traditionally, the right to name a new element goes to the first to find it, but it is not clear that the world would accept that premise if the work is done secretly.)'' \cite{2000Hof01}.

Ghiorso et al. succeeded and submitted the first observation of the einsteinium isotope $^{246}$Es in November 1953 \cite{1954Ghi01}. They did not want this observation to be regarded as the discovery of einsteinium and added the note: ``There is unpublished information relevant to element 99 at the University of California, Argonne National Laboratory, and Los Alamos Scientific Laboratory. Until this information is published the question of the first preparation should not be prejudged on the basis of this paper.'' The same statement was included the paper reporting the observation of $^{253}$Es submitted a month later (December 1953) \cite{1954Tho01}. The first identification of $^{254}$Es (February 1954 \cite{1954Fie01}) and $^{255}$Es (March 1954 \cite{1954Cho01}) were submitted also prior to the official announcement of the discovery of the new element. This announcement was finally made in the summer of 1955 with the publication of the article ``New Elements Einsteinium and Fermium, Atomic Numbers 99 and 100'' \cite{1955Ghi01}.

The early accounts of the events were not specific about the details: ``Without going into the details, it may be pointed out that such experiments involving the groups at the three laboratories led to the positive identification of isotopes of elements 99 and 100'' \cite{1958Sea01,1963Sea01,1990Sea01}. Only later were the difficult discussions regarding the publication strategy between the research groups involved described in detail \cite{2000Hof01}. It is interesting to note that the loss of life during the collection of samples from the thermonuclear explosion was only mentioned in the more recent accounts \cite{1978Sea01,1990Sea01} of the discovery of these transuranium elements: ``These samples cost the life of First Lieutenant Jimmy Robinson, who waited too long before he went home, tried to land on Eniwetok, and ditched about a mile short of the runway'' \cite{1978Sea01}.

In the discovery paper the authors suggested to name the new element with Z = 99 Einsteinium with the symbol ``E''. The International Union of Pure and Applied Chemistry (IUPAC) adopted the name but changed to symbol to ``Es'' at the 19$^{th}$ IUPAC Conference in Paris 1957 \cite{2005Kop01,1957IUP01}.

\section{Discovery of $^{241-257}$Es}

Seventeen einsteinium isotopes from A = $241-257$ have been discovered so far; there are no stable einsteinium isotopes. According to the HFB-14 model \cite{2007Gor01}, einsteinium isotopes ranging from $^{235}$Es through $^{328}$Es plus $^{330}$Es and $^{332}$Es should be particle stable. Thus, there remain about 80 isotopes to be discovered. In addition, it is estimated that 16 additional nuclei beyond the proton dripline could live long enough to be observed \cite{2004Tho01}. Less than 20\% of all possible einsteinium isotopes have been produced and identified so far.

\begin{figure}
 	\centering
	\includegraphics[width=12cm]{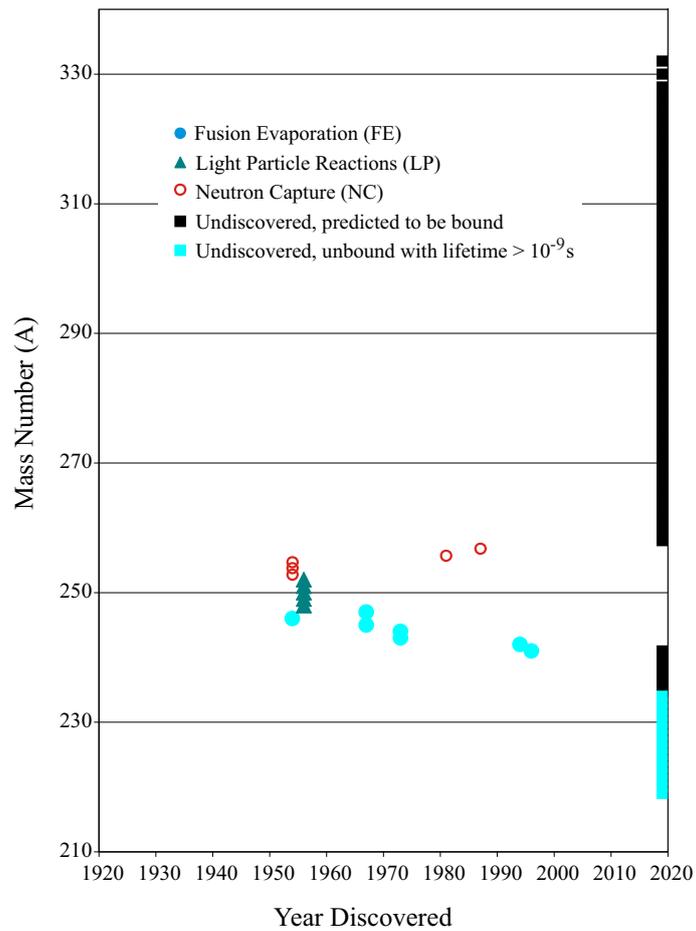}
	\caption{Einsteinium isotopes as a function of when they were discovered. The different production methods are indicated. The solid black squares on the right hand side of the plot are isotopes predicted to be bound by the HFB-14 model.  On the proton-rich side the light blue squares correspond to unbound isotopes predicted to have lifetimes larger than $\sim 10^{-9}$~s.}
	\label{f:year-es}
\end{figure}

Figure \ref{f:year-es} summarizes the year of first discovery for all einsteinium isotopes identified by the method of discovery.  The range of isotopes predicted to exist is indicated on the right side of the figure.  Only three different reaction types were used to produce the radioactive einsteinium isotopes; heavy-ion fusion evaporation (FE), light-particle reactions (LP), and neutron-capture reactions (NC). Heavy ions are all nuclei with an atomic mass larger than A = 4 \cite{1977Gru01}. In the following, the discovery of each einsteinium isotope is discussed in detail and a summary is presented in Table 1.

\subsection{$^{241}$Es}\vspace{0.0cm}
In 1996, Ninov et al. discovered $^{241}$Es at the Gesellschaft f\"{u}r Schwerionenforschung (GSI) in Darmstadt, Germany, as reported in the paper ``Identification of new mendelevium and einsteinium isotopes in bombardments of $^{209}$Bi with $^{40}$Ar'' \cite{1996Nin01}. $^{40}$Ar beams were accelerated to 4.78, 4.93, and 5.12 A$\cdot$MeV with the UNILAC accelerator and bombarded $^{209}$Bi targets. $^{241}$Es was created via the $\alpha$-decay of $^{245}$Md which was produced in the fusion-evaporation reaction $^{209}$Bi($^{40}$Ar,4n). The time and position correlation of $\alpha$-particles with evaporation residues were measured. ``Since the chosen bombarding energy further coincided with the maxiumum of the 4n deexcitation channel, we assign this decay sequence to $^{245}$Md and its daughter $^{241}$Es.'' The half-life of 8$^{+6}_{-4}$~s is currently the only measured value for $^{241}$Es.

\subsection{$^{242}$Es}\vspace{0.0cm}
The discovery of $^{242}$Es was reported in 1994 by Lazarev et al. in ``Cross sections of the (HI,$\alpha$n) channel in the cold-fusion-type reactions $^{209}$Bi + $^{40}$Ar and $^{208}$Pb + $^{37}$Cl'' \cite{1994Laz01}. The isotope was identified using the catcher technique with off-line chemical separation following the irradiation of the targets with beams from the U400 cyclotron of the JINR Laboratory of Nuclear Reactions at Dubna. ``The most probable explanation of this activity seems to be the EC-delayed fission in the decay chain $^{242}$Es $\rightarrow ^{242}$Cf.'' The extracted half-life is close to the accepted vales of  13.5(25)~s. It should be mentioned that the observation of $^{242}$Es had been reported nine years earlier in an unpublished report \cite{1985Hin01}. Ninov et al. \cite{1996Nin01} does not refer to the work by Lazarev et al. in their 1996 paper about the production of the new neutron deficient isotopes $^{241}$Es and $^{242}$Es.

\subsection{$^{243,244}$Es}\vspace{0.0cm}
In 1973, Eskola et al. reported the discovery of $^{243}$Es and $^{244}$Es in ``Two new isotopes of einsteinium, $^{243}$Es and $^{244}$Es'' \cite{1973Esk01}. A $^{233}$U target was bombarded with $^{15}$N ions from the Berkeley HILAC and the isotopes were produced in the fusion evaporation reactions $^{233}$U($^{15}$N,5n)$^{243}$Es and $^{233}$U($^{15}$N,4n)$^{244}$Es. The isotopes were identified from the excitation functions, and $\alpha$-spectra were measured. ``The $\alpha$-particle group at 7.89~MeV has been assigned to $^{243}$Es and that at 7.57~MeV to $^{244}$Es.'' The measured half-lives of 21(2)~s and 37(4)~s for $^{243}$Es and $^{244}$Es, respectively, correspond to the currently accepted values.

\subsection{$^{245}$Es}\vspace{0.0cm}
The discovery of $^{245}$Es was first published by Mikheev et al. in 1967 in ``Synthesis of Einsteinium Isotopes in Reactions with Nitrogen Ions'' \cite{1967Mik01}. $^{245}$Es was produced in the fusion-evaporation reaction $^{238}$U($^{14}$N,7n)$^{245}$Es following the acceleration of the $^{14}$N ions by the JINR cyclotron in Dubna. The recoil atoms were transported in front of charged-particle detectors with a helium jet. ``The half-life 1.33$\pm$0.15 min, obtained by the maximum-likelihood method, is also in good agreement with the published data on Es$^{245}$.'' The measured half-life agrees with the currently accepted valus of 1.1(1)~m. The ``published'' data that Mikheev et al. refer to are unpublished results that were subsequently quoted in a review \cite{1961Ghi01}.

\subsection{$^{246}$Es}\vspace{0.0cm}
Ghiorso et al. discovered $^{246}$Es in 1954 as reported in ``Reactions of U$^{238}$ with Cyclotron-Produced Nitrogen Ions'' \cite{1954Ghi01}. The Berkeley Crocker Laboratory 60-inch cyclotron produced $^{14}$N beams that bombarded $^{238}$U targets and $^{246}$Es was produced in the fusion-evaporation reaction $^{238}$U($^{14}$N,6n). Electron-capture and $\alpha$-particles were detected following chemical separation. $^{246}$Es was ``observed only through growth of its 1.5-day Cf$^{246}$ daughter.'' Electron-capture with a half-life of ``minutes'' was measured. The currently accepted half-life is 7.7(5)~min. The paper was the first unclassified publication of an einsteinium isotope. However, the paper states in a footnote that ``There is unpublished information relevant to element 99 at the University of California, Argonne National Laboratory, and Los Alamos Scientific Laboratory. Until this information is published the question of the first preparation should not be prejudged on the basis of this paper.'' The official announcement of the discovery of the new element with Z = 99 was made in 1955 \cite{1955Ghi01}. $^{253}$Es was identified on December 19-20, 1952 from uranium which had been irradiated by neutrons in the ``Mike'' thermonuclear explosion in November 1952.

\subsection{$^{247}$Es}\vspace{0.0cm}
$^{247}$Es was identified for the first time by Mikheev et al. in 1967 in the same paper in which they reported the first observation of $^{245}$Es: ``Synthesis of Einsteinium Isotopes in Reactions with Nitrogen Ions'' \cite{1967Mik01}. $^{247}$Es was produced in the fusion-evaporation reaction $^{238}$U($^{14}$N,5n)$^{247}$Es following the acceleration of the $^{14}$N ions by the JINR cyclotron in Dubna. The recoil atoms were transported in front of charged-particle detectors with a helium jet. ``Our results show that the isotopes Es$^{247}$ and Es$^{246}$ have practically identical energies of the main $\alpha$-particle groups, 7.33$\pm$0.03 MeV, and nearly equal half-lives, 5.0$\pm$0.3 and 7.7$\pm$0.5 min.'' The measured half-life agrees with the currently accepted value of 4.55(26)~m. In 1954 Ghiorso et al. \cite{1954Ghi01} had observed a 7.3~m half-live which was tentatively assigned to $^{247}$Es, where the mass identification was based only on nuclear systematics. Thus it is probably that the observed half-life corresponded to the decay of $^{246}$Es.

\subsection{$^{248}$Es}\vspace{0.0cm}
Chetham-Strode and Hold reported the first observation of $^{248}$Es in 1956 in ``New Isotope Einsteinium-248'' \cite{1956Che01}. Deuterons between 18 and 22 MeV were accelerated by the 60-inch cyclotron of the Berkeley Crocker Laboratory and bombarded a target containing 10$^{13}$ atoms of monoisotopic $^{249}$Cf. $^{248}$Es was produced in the reaction $^{249}$Cf(d,3n) and identified on the basis of partial excitation function. ``A new isotope of einsteinium, E$^{248}$, has been identified among the products of the bombardment of Cf$^{249}$ with 18- to 22-Mev deuterons. It decays principally by electron capture with a half-life of 25$\pm$5 minutes and also by the emission of (6.87$\pm$0.02)-Mev alpha particles.'' The measured half-life is included in the average for the currently accepted value of 27(5)~m.

\subsection{$^{249-252}$Es}\vspace{0.0cm}
The isotopes $^{249}$Es, $^{250}$Es, $^{251}$Es, and $^{252}$Es were produced by Harvey et al. in 1956 as reported in ``New Isotopes of Einsteinium'' \cite{1956Har01}. A gold foil deposited with 3$\times$10$^{13}$ atoms of $^{249}$Bk was bombarded with 20$-$40 MeV $\alpha$-particles accelerated by the Berkeley Crocker Laboratory 60-inch cyclotron: ``... the 280-day $\beta^-$-emitter Bk$^{249}$ was bombarded with helium ions from 20 to 40 Mev, and reactions of the type ($\alpha$,xn) were studied radiochemically. Such reactions can produce four previously unobserved isotopes of einsteinium (symbol E, atomic number 99) with mass numbers from 249 to 252.''  The isotopes were chemically separated and their decay via electron capture and/or $\alpha$-particles measured. The quoted half-lives of 2~h ($^{249}$Es), 8~h ($^{250}$Es), and 1.5~d ($^{251}$Es), agree with the presently accepted values of 102.2(6)~m, 8.6(1)~h, and 33(1)~h, respectively. Although the estimated half-life of $\sim$140~d for $^{252}$Es is significantly shorter than the adapted value of 471.7(19)~d we still credit Harvey et al. with the discovery because the measured $\alpha$-particle energy (6.64~MeV) agrees with later measurements \cite{1973Fie01}.

\subsection{$^{253}$Es}\vspace{0.0cm}
Thompson et al. from Berkeley were the first to publish unclassified evidence for the existence of $^{253}$Es in 1954 in their paper ``Transcurium Isotopes Produced in the Neutron Irradiation of Plutonium'' \cite{1954Tho01}. $^{239}$Pu was irradiated with neutrons in the Idaho Materials Testing Reactor and $^{253}$Cf was chemically separated. ``The isotope of element 99 emitting 6.6-MeV alpha particles is logically assigned as 99$^{253}$. A reasonable half-life estimated from systematics, assuming a hindrance factor of ten, would be very roughly a month.'' The currently accepted half-life is 20.47(3)~d. The paper was published shortly after the report on the discovery of $^{246}$Es and contained the same statement regarding the discovery of element 99: ``There is unpublished information relevant to element 99 at the University of California, Argonne National Laboratory, and Los Alamos Scientific Laboratory. Until this information is published the question of the first preparation should not be prejudged on the basis of this paper.'' The Argonne group \cite{1954Stu01} published their results on $^{253}$Es only a few weeks later, stating ``A number of arguments indicate that the probable mass assignment of the element-99 isotope is 253.'' The paper included a footnote with regards to the discovery of the new element: ``These elements (99 and 100) have previously been discovered in other work at Argonne National Laboratory, University of California Radiation Laboratory, and Los Alamos Scientific Laboratory, not yet published.'' $^{253}$Es was the einsteinium isotope credited with the discovery of element 99, published in 1955 \cite{1955Ghi01}. It was identified on December 19-20, 1952 from uranium which had been irradiated by neutrons in the ``Mike'' thermonuclear explosion in November 1952.

\subsection{$^{254}$Es}\vspace{0.0cm}
The observation of $^{254}$Es was reported in 1954 by Fields et al. from Argonne in the paper ``Additional Properties of Isotopes of Elements 99 and 100'' \cite{1954Fie01}. ``This note describes the results obtained from a four-day irradiation of an element-99 fraction with californium impurity in the Materials Testing Reactor (MTR) at Arco, Idaho to produce the following reaction: 99$^{253}$(n,$\gamma$)99$^{254} \overset{\beta}{\rightarrow} 100^{254}$.'' The half-life was measured to be 37(1)~h and corresponds to an isomer. A few weeks earlier the Berkeley group had described the formation of 100$^{254}$ and the ``... possible reaction sequence leading to its production might be the following ...'' included $^{254}$Es \cite{1954Har01}. However, no properties of $^{254}$Es were reported. The Berkeley group confirmed the half-life measurement \cite{1954Cho01} only a few weeks following the Argonne paper.

\subsection{$^{255}$Es}\vspace{0.0cm}
Choppin et al. from Berkeley reported the first observation of $^{255}$Es in 1954 in their paper ``Nuclear Properties of Some Isotopes of Californium, Elements 99 and 100'' \cite{1954Cho01}. The isotope was formed in intense neutron bombardments of plutonium and high-mass nuclides. ``The 7.1-Mev alpha activity was also observed in purified element-99 fractions, and thus it was deduced that is also has a $\beta^-$ emitting parent. Rough decay measurements of this activity in the element-99 fraction indicate that the half-life of this parent, presumably 99$^{255}$, is approximately one month.'' The presently adapted half-life is 39.8(12)~d.

\subsection{$^{256}$Es}\vspace{0.0cm}
In 1981 Lougheed et al. from Lawrence Livermore National Laboratory discovered $^{256}$Es as reported in ``Two new isotopes with N = 157: $^{256}$Es and $^{255}$Cf'' \cite{1981Lou01}. A beryllium foil containing $^{255}$Es was irradiated by neutrons in the core of the Livermore Pool-Type Reactor and einsteinium was then chemically separated: ``... we have identified ... $^{256}$Es with a half life of 25$\pm$2.4 min. Our identification is based on chemical identification of the elements and on the known nuclear properties of their daughters.'' This half-life is currently the only measurement of $^{256}$Es. Indirect evidence for the existence of $^{256}$Es had been shown in 1955 in the production of $^{256}$Fm by neutron capture on $^{255}$Es \cite{1955Cho01}.

\subsection{$^{257}$Es}\vspace{0.0cm}
$^{257}$Es was discovered in 1987 by Popov et al. from Dimitrovgrad, Russia, and described in the paper ``Determining the half-lives of $^{253}$Es, $^{254}$Es, $^{254m}$Es, $^{255}$Es, $^{257}$Es, $^{256}$Fm'' \cite{1987Pop01}. $^{257}$Es was produced by neutron irradiation of $^{252}$Cf targets in the high-flux SM-2 reactor of the Research Institute of Nuclear Reactors and identified following chemical separation using X-ray, $\gamma$- and $\alpha$-spectrometry. ``The gamma lines at 6.5, 7.3, 25.7, 45.2, 46.0 and 49.0 keV evidently accompany the beta decay of $^{257}$Es. Here T$_\beta$ for $^{257}$Es is found as 7.7$\pm$0.2~d.'' This is currently the only half-life measurement of $^{257}$Es.

\section{Summary}
The discoveries of the known scandium, titanium, mercury, and einsteinium isotopes have been compiled and the methods of their production discussed.

The limit for observing long lived scandium isotopes beyond the proton dripline which can be measured by implantation decay studies has most likely been reached with the discovery of $^{40}$Sc and the observation that of $^{39}$Sc is unbound with respect to proton emission by 580 keV. The discovery of especially the light scandium isotopes was difficult. Five isotopes - two of twice - were initially identified incorrectly ($^{40-42}$Sc, $^{44}$Sc and $^{47}$Sc). The half-life of $^{49}$Sc had first been assigned to $^{44}$Sc and then to $^{41}$Sc.

The limit for observing long lived titanium isotopes beyond the proton dripline which can be measured by implantation decay studies has most likely been reached with the discovery of $^{39}$Ti and the non-observation of $^{38}$Ti. The discovery of the titanium isotopes was straight forward. Only the half-life of $^{51}$Ti was initially incorrect.

The identification of the odd mercury isotopes between mass 183 and 205 proofed to be especially difficult. $^{183}$Hg, $^{197}$Hg, $^{203}$Hg, and $^{205}$Hg, were initially incorrectly identified. The half-life of $^{203}$Hg had been observed previously but without a mass assignment. For $^{189}$Hg and $^{191}$Hg it could not be determined if the ground state or an isomeric state was observed, while for $^{193}$Hg and $^{195}$Hg the isomeric state was most likely observed first. The half-life measurement of $^{194}$Hg was initially reported incorrect. It is interesting to note that the half-life of $^{207}$Hg had been reported in a 1981 conference proceedings and even after over 25 years has not been published in a refereed journal. Finally, the discovery of $^{209}$Hg was reported by the authors simultaneously in two different journals.

The discovery of the element einsteinium is credited to A. Ghiorso et al., who observed $^{253}$Es in uranium which had been irradiated by neutrons in the ``Mike'' thermonuclear explosion in November 1952 \cite{1955Ghi01}. However, this classified work was not published until 1955. In the meantime several groups succeeded in producing einsteinium isotopes in fusion-evaporation reactions ($^{246}$Es) and neutron-capture reactions ($^{253}$Es). These publications always are careful to point out that their observations should not be considered as the discovery of einsteinium. Nevertheless, following our guidelines to acknowledge first publications in refereed journals we credit these papers with the first observation of the specific isotope. This should not be interpreted as the discovery of the element Einsteinium which is accepted to have been discovered on December 19-20, 1952 from uranium irradiated in the ``Mike'' thermonuclear explosion.

Two einsteinium isotopes were discovered years prior to the first refereed publications; however, the results had only been presented in internal reports ($^{242}$Es) or unpublished work quoted in a review article ($^{245}$Es).

\ack

The main research on the individual elements were performed by DM (scandium, titanium, and mercury) and AB (einsteinium). AB acknowledges the support of the High School Honors Science Program at Michigan State University and would like to thank A. Fritsch, M. Heim, A. Schuh, and A. Shore for help during the project. This work was supported by the National Science Foundation under grant No. PHY06-06007 (NSCL).

%%% Here we use thebibliography environment to produce the reference list,
%%% but you can use BibTeX as well:
\bibliography{../isotope-discovery-references}

\newpage

%%% Please start a new page by uncommenting the next
\newpage

\TableExplanation

\bigskip
\renewcommand{\arraystretch}{1.0}

\section{Table 1.\label{tbl1te} Discovery of scandium, titanium, mercury, and einsteinium isotopes }
\begin{tabular*}{0.95\textwidth}{@{}@{\extracolsep{\fill}}lp{5.5in}@{}}
\multicolumn{2}{p{0.95\textwidth}}{ }\\

Isotope & Scandium, titanium, mercury, and einsteinium isotope \\
Author & First author of refereed publication \\
Journal & Journal of publication \\
Ref. & Reference \\
Method & Production method used in the discovery: \\

    & FE: fusion evaporation \\
    & LP: light-particle reactions (including neutrons) \\
    & TR: heavy-ion transfer reactions \\
    & PI: pion induced reactions \\
    & MS: mass spectroscopy \\
    & AD: alpha decay \\
    & NC: neutron capture reactions \\
    & DI: deep-inelastic reactions \\
    & SP: spallation \\
    & PF: projectile fragmentation or fission \\

Laboratory & Laboratory where the experiment was performed\\
Country & Country of laboratory\\
Year & Year of discovery \\
\end{tabular*}
\label{tableI}

\datatables % This command is necessary to get the table names in toc

%% One-page data tables are also best formatted using the longtable
%% environment:
%\begin{longtable}{c}
%\caption{This is the First Data Table}\\
%\endhead\\
%\end{longtable}

%% If the table is to span over the whole text width, we set:

\setlength{\LTleft}{0pt}
\setlength{\LTright}{0pt}

% To avoid ``Overfull \hboxes...'' decrease the intercolumn spacing:

\setlength{\tabcolsep}{0.5\tabcolsep}

\renewcommand{\arraystretch}{1.0}

\footnotesize % we need to squeeze the font size a lot!

\begin{longtable}{@{\extracolsep\fill}llllllll@{}}
\caption{Discovery of Scandium Titanium, Mercury, and Einsteinium Isotopes. See page\ \pageref{tbl1te} for Explanation of Tables}
Isotope & Author & Journal & Ref. & Method & Laboratory & Country & Year\\
\hline\\
\endfirsthead\\
\caption[]{(continued)}
Isotope & Author & Journal & Ref. & Method & Laboratory & Country & Year\\
\hline\\
\endhead
$^{39}$Sc  & C.L. Woods & Nucl. Phys. A &\cite{1988Woo01}& TR & Canberra & Australia &1988 \\
$^{40}$Sc  & N.W. Glass & Phys. Rev. &\cite{1955Gla01}& LP & UCLA & USA &1955 \\
$^{41}$Sc  & D.R. Elliott & Phys. Rev. &\cite{1941Ell01}& LP & Purdue & USA &1941 \\
$^{42}$Sc  & H. Morinaga & Phys. Rev. &\cite{1955Mor01}& LP & Purdue & USA &1955 \\
$^{43}$Sc  & O.R. Frisch & Nature &\cite{1935Fri01}& LP & Copenhagen & Denmark &1935 \\
$^{44}$Sc  & H. Walke & Phys. Rev. &\cite{1937Wal01}& LP & Berkeley & USA &1937 \\
$^{45}$Sc  & F.W. Aston & Nature &\cite{1923Ast01}& MS & Cambridge & UK &1923 \\
$^{46}$Sc  & G. Hevesy & Mat.-fys. Medd. &\cite{1936Hev01}& NC & Copenhagen & Denmark &1936 \\
$^{47}$Sc  & C.T. Hibdon & Phys. Rev. &\cite{1945Hib01}& LP & Ohio State & USA &1945 \\
$^{48}$Sc  & H. Walke & Phys. Rev. &\cite{1937Wal03}& LP & Berkeley & USA &1937 \\
$^{49}$Sc  & H. Walke & Phys. Rev. &\cite{1940Wal01}& LP & Berkeley & USA &1940 \\
$^{50}$Sc  & A. Poularikas & Phys. Rev. &\cite{1959Pou01}& LP & Arkansas & USA &1959 \\
$^{51}$Sc  & J.R. Erskine & Phys. Rev. &\cite{1966Ers01}& LP & Argonne & USA &1966 \\
$^{52}$Sc  & H. Breuer & Phys. Rev. C &\cite{1980Bre01}& DI & Berkeley & USA &1980 \\
$^{53}$Sc  & H. Breuer & Phys. Rev. C &\cite{1980Bre01}& DI & Berkeley & USA &1980 \\
$^{54}$Sc  & X.L. Tu & Z. Phys. A &\cite{1990Tu01}& SP & Los Alamos & USA &1990 \\
$^{55}$Sc  & X.L. Tu & Z. Phys. A &\cite{1990Tu01}& SP & Los Alamos & USA &1990 \\
$^{56}$Sc  & M. Bernas & Phys. Lett. B &\cite{1997Ber01}& PF & Darmstadt & Germany &1997 \\
$^{57}$Sc  & M. Bernas & Phys. Lett. B &\cite{1997Ber01}& PF & Darmstadt & Germany &1997 \\
$^{58}$Sc  & M. Bernas & Phys. Lett. B &\cite{1997Ber01}& PF & Darmstadt & Germany &1997 \\
$^{59}$Sc  & O.B. Tarasov & Phys. Rev. Lett. &\cite{2009Tar01}& PF & Michigan State & USA &2009 \\
$^{60}$Sc  & O.B. Tarasov & Phys. Rev. Lett. &\cite{2009Tar01}& PF & Michigan State & USA &2009 \\
$^{61}$Sc  & O.B. Tarasov & Phys. Rev. Lett. &\cite{2009Tar01}& PF & Michigan State & USA &2009 \\
&  &  & &  &  & & \\
&  &  & &  &  & & \\
$^{39}$Ti   & C. Detraz & Nucl. Phys. A &\cite{1990Det01}& PF & GANIL & France &1990 \\
$^{40}$Ti   & C.L. Morris & Phys. Rev. C &\cite{1982Mor01}& PI & Los Alamos & USA &1982 \\
$^{41}$Ti   & P.L. Reeder& Phys. Rev. Lett. &\cite{1964Ree01}& LP & Brookhaven & USA &1964 \\
$^{42}$Ti   & H.C. Bryant & Nucl. Phys. &\cite{1964Bry01}& LP & Los Alamos & USA &1964 \\
$^{43}$Ti   & A.D. Schelberg & Rev. Sci. Instrum. &\cite{1948Sch01}& LP & Indiana & USA &1948 \\
$^{44}$Ti   & R.A. Sharp & Phys. Rev. &\cite{1954Sha01}& LP & Harvard & USA &1954 \\
$^{45}$Ti   & J.S.V. Allen & Phys. Rev. &\cite{1941All01}& LP & Ohio State & USA &1941 \\
$^{46}$Ti   & F.W. Aston & Nature &\cite{1934Ast02}& MS & Cambridge & UK &1934 \\
$^{47}$Ti   & F.W. Aston & Nature &\cite{1934Ast02}& MS & Cambridge & UK &1934 \\
$^{48}$Ti   & F.W. Aston & Nature &\cite{1923Ast01}& MS & Cambridge & UK &1923 \\
$^{49}$Ti   & F.W. Aston & Nature &\cite{1934Ast02}& MS & Cambridge & UK &1934 \\
$^{50}$Ti   & F.W. Aston & Nature &\cite{1934Ast02}& MS & Cambridge & UK &1934 \\
$^{51}$Ti   & L. Seren& Phys. Rev. &\cite{1947Ser01}& NC & Argonne & USA &1947 \\
$^{52}$Ti   & D.C. Williams & Phys. Lett. &\cite{1966Wil01}& LP & Los Alamos & USA &1966 \\
$^{53}$Ti   & L.A. Parks & Phys. Rev. C &\cite{1977Par01}& FE & Argonne & USA &1977 \\
$^{54}$Ti   & D. Guerreau & Z. Phys. A &\cite{1980Gue01}& DI & Orsay & France &1980 \\
$^{55}$Ti   & D. Guerreau & Z. Phys. A &\cite{1980Gue01}& DI & Orsay & France &1980 \\
$^{56}$Ti   & H. Breuer & Phys. Rev. C &\cite{1980Bre01}& DI & Berkeley & USA &1980 \\
$^{57}$Ti   & D. Guillemaud-Mueller & Z. Phys. A &\cite{1985Gui01}& PF & GANIL & France &1985 \\
$^{58}$Ti   & M. Weber & Z. Phys. A &\cite{1992Web01}& PF & Darmstadt & Germany &1992 \\
$^{59}$Ti   & M. Bernas & Phys. Lett. B &\cite{1997Ber01}& PF & Darmstadt & Germany &1997 \\
$^{60}$Ti   & M. Bernas & Phys. Lett. B &\cite{1997Ber01}& PF & Darmstadt & Germany &1997 \\
$^{61}$Ti   & M. Bernas & Phys. Lett. B &\cite{1997Ber01}& PF & Darmstadt & Germany &1997 \\
$^{62}$Ti   & O.B. Tarasov & Phys. Rev. Lett. &\cite{2009Tar01}& PF & Michigan State & USA &2009 \\
$^{63}$Ti   & O.B. Tarasov & Phys. Rev. Lett. &\cite{2009Tar01}& PF & Michigan State & USA &2009 \\
&  &  & &  &  & & \\
&  &  & &  &  & & \\
$^{171}$Hg & H. Kettunen & Phys. Rev. C &\cite{2004Ket01}& FE & Jyvaskyla & Finland &2004 \\
$^{172}$Hg & D. Seweryniak & Phys. Rev. C &\cite{1999Sew01}& FE & Argonne & USA &1999 \\
$^{173}$Hg & D. Seweryniak & Phys. Rev. C &\cite{1999Sew01}& FE & Argonne & USA &1999 \\
$^{174}$Hg & J. Uusitalo & Z. Phys. A &\cite{1997Uus01}& FE & Jyvaskyla & Finland &1997 \\
$^{175}$Hg & J.R.H. Schneider & Z. Phys. A &\cite{1983Sch01}& FE & Darmstadt & Germany &1983 \\
$^{176}$Hg & J.R.H. Schneider & Z. Phys. A &\cite{1983Sch01}& FE & Darmstadt & Germany &1983 \\
$^{177}$Hg & C. Cabot & Compt. Rend. Acad. Sci. &\cite{1975Cab02}& FE & Orsay & France &1975 \\
$^{178}$Hg & P.G. Hansen & Nucl. Phys. A &\cite{1971Han01}& SP & CERN & Switzerland &1971 \\
$^{179}$Hg & P.G. Hansen & Nucl. Phys. A &\cite{1970Han01}& SP & CERN & Switzerland &1970 \\
$^{180}$Hg & P.G. Hansen & Nucl. Phys. A &\cite{1970Han01}& SP & CERN & Switzerland &1970 \\
$^{181}$Hg & P.G. Hansen & Phys. Lett. B &\cite{1969Han01}& SP & CERN & Switzerland &1969 \\
$^{182}$Hg & A.G. Demin & Nucl. Phys. A &\cite{1968Dem01}& FE & Dubna & Russia &1968 \\
$^{183}$Hg & P.G. Hansen & Phys. Lett. B &\cite{1969Han01}& SP & CERN & Switzerland &1969 \\
$^{184}$Hg & P.G. Hansen & Phys. Lett. B &\cite{1969Han01}& SP & CERN & Switzerland &1969 \\
$^{185}$Hg & G. Albouy & J. Phys. Radium &\cite{1960Alb01}& SP & Orsay & France &1960 \\
$^{186}$Hg & G. Albouy & J. Phys. Radium &\cite{1960Alb01}& SP & Orsay & France &1960 \\
$^{187}$Hg & G. Albouy & J. Phys. Radium &\cite{1960Alb01}& SP & Orsay & France &1960 \\
$^{188}$Hg & N. Poffe & J. Phys. Radium &\cite{1960Pof01}& SP & Orsay & France &1960 \\
$^{189}$Hg & W.G. Smith & Phys. Rev. &\cite{1955Smi01}& LP & Berkeley & USA &1955 \\
$^{190}$Hg & G. Albouy & Compt. Rend. Acad. Sci. &\cite{1959Alb01}& SP & Orsay & France &1959 \\
$^{191}$Hg & L.P. Gillon & Phys. Rev. &\cite{1954Gil01}& LP & Harvard & USA &1954 \\
$^{192}$Hg & R.W. Fink & J. Am. Chem. Soc. &\cite{1952Fin01}& LP & Rochester & USA &1952 \\
$^{193}$Hg & R.W. Fink & J. Am. Chem. Soc. &\cite{1952Fin01}& LP & Rochester & USA &1952 \\
$^{194}$Hg & W.J. Tomlinson III & Phys. Rev. Lett. &\cite{1962Tom01}& LP & MIT & USA &1962 \\
$^{195}$Hg & R.W. Fink & J. Am. Chem. Soc. &\cite{1952Fin01}& LP & Rochester & USA &1952 \\
$^{196}$Hg & F.W. Aston & Proc. Roy. Soc. &\cite{1930Ast01}& MS & Cambridge & UK &1930 \\
$^{197}$Hg & R. Sherr & Phys. Rev. &\cite{1941She01}& LP & Harvard & USA &1941 \\
$^{198}$Hg & F.W. Aston & Nature &\cite{1925Ast01}& MS & Cambridge & UK &1925 \\
$^{199}$Hg & F.W. Aston & Nature &\cite{1925Ast01}& MS & Cambridge & UK &1925 \\
$^{200}$Hg & F.W. Aston & Nature &\cite{1925Ast01}& MS & Cambridge & UK &1925 \\
$^{201}$Hg & F.W. Aston & Nature &\cite{1925Ast01}& MS & Cambridge & UK &1925 \\
$^{202}$Hg & F.W. Aston & Nature &\cite{1920Ast01}& MS & Cambridge & UK &1920 \\
$^{203}$Hg & G. Friedlander & Phys. Rev. &\cite{1943Fri01}& NC & Berkeley & USA &1943 \\
$^{204}$Hg & F.W. Aston & Nature &\cite{1920Ast01}& MS & Cambridge & UK &1920 \\
$^{205}$Hg & R.S. Krishnan & Proc. Camb. Phil. Soc. &\cite{1940Kri01}& LP & Cambridge & UK &1940 \\
$^{206}$Hg & M.J. Nurmia & Nature &\cite{1961Nur01}& AD & Helsinki & Finland &1961 \\
$^{207}$Hg & S. Mirzadeh & Radiochem. Radioanal. Lett. &\cite{1982Mir01}& LP & Brookhaven & USA &1982 \\
$^{208}$Hg & L. Zhang & Phys. Rev. C &\cite{1994Zha01}& LP & Lanzhou & China &1994 \\
$^{209}$Hg & L. Zhang & Phys. Rev. C &\cite{1998Zha01}& LP & Lanzhou & China &1998 \\
$^{210}$Hg & M. Pfuetzner & Phys. Lett. B &\cite{1998Pfu01}& PF & Darmstadt & Germany &1998 \\
&  &  & &  &  & & \\
&  &  & &  &  & & \\
$^{241}$Es & V. Ninov & Z. Phys. A &\cite{1996Nin01}& FE & Darmstadt & Germany &1996 \\
$^{242}$Es & Yu.A. Lazarev & Nucl. Phys. A &\cite{1994Laz01}& FE & Dubna & Russia &1994 \\
$^{243}$Es & P. Eskola & Physica Fennica &\cite{1973Esk01}& FE & Berkeley & USA &1973 \\
$^{244}$Es & P. Eskola & Physica Fennica &\cite{1973Esk01}& FE & Berkeley & USA &1973 \\
$^{245}$Es & V.L. Mikheev & Sov. J. Nucl. Phys. &\cite{1967Mik01}& FE & Dubna & Russia &1967 \\
$^{246}$Es & A. Ghiorso & Phys. Rev. &\cite{1954Ghi01}& FE & Berkeley & USA &1954 \\
$^{247}$Es & V.L. Mikheev & Sov. J. Nucl. Phys. &\cite{1967Mik01}& FE & Dubna & Russia &1967 \\
$^{248}$Es & A. Chetham & Phys. Rev. &\cite{1956Che01}& LP & Berkeley & USA &1956 \\
$^{249}$Es & B.G. Harvey & Phys. Rev. &\cite{1956Har01}& LP & Berkeley & USA &1956 \\
$^{250}$Es & B.G. Harvey & Phys. Rev. &\cite{1956Har01}& LP & Berkeley & USA &1956 \\
$^{251}$Es & B.G. Harvey & Phys. Rev. &\cite{1956Har01}& LP & Berkeley & USA &1956 \\
$^{252}$Es & B.G. Harvey & Phys. Rev. &\cite{1956Har01}& LP & Berkeley & USA &1956 \\
$^{253}$Es & S.G. Thompson & Phys. Rev. &\cite{1954Tho01}& NC & Berkeley & USA &1954 \\
$^{254}$Es & P.R. Fields & Phys. Rev. &\cite{1954Fie01}& NC & Argonne & USA &1954 \\
$^{255}$Es & G.R. Choppin & Phys. Rev. &\cite{1954Cho01}& NC & Berkeley & USA &1954 \\
$^{256}$Es & R.W. Lougheed & J. Inorg. Nucl. Chem. &\cite{1981Lou01}& NC & Livermore & USA &1981 \\
$^{257}$Es & Yu.S. Popov & Sov. J. Radiochem. &\cite{1987Pop01}& NC & Dimitrovgrad & Russia &1987 \\
 \\
\end{longtable}

\end{document}